\documentclass[a4paper,12pt]{article}
\usepackage{graphicx}
\usepackage{epsfig}
\usepackage{jcappub}

\begin{document}

\title{DBI Galileon and Late time acceleration of the universe}
\author{Sampurnanand $^{1,2}$, Anjan A Sen $^1$}
\affiliation{$^1$Center For Theoretical Physics, Jamia Millia Islamia, New Delhi-110025, India}
\affiliation{$^2$Department of Physics and Astrophysics, University of Delhi, Delhi-110007, India}

\emailAdd{sampurna@physics.du.ac.in,  aasen@jmi.ac.in}

\abstract{
We consider $1+3$  dimensional maximally symmetric Minkowski brane embedded in a $1+4$ dimensional maximally symmetric Minkowski background. The resulting $1+3$ dimensional effective field theory is of DBI (Dirac-Born-Infeld) Galileon type. We use this model to study the late time acceleration of the universe. We study the deviation of the model from the concordance $\Lambda$CDM behaviour. Finally we put constraints on the model parameters using various observational data.}
\date{\today}

\maketitle

\section{Introduction}
Providing a satisfactory explaination of the late time acceleration of the universe is one of the most challenging tasks for cosmologists and particle physicists at present \cite{review}. The most studied approach invloves adding an exotic form of energy with negative pressure (termed as dark energy) in the energy budget of the universe. The cosmological constant  ( with $w= \frac{p}{\rho}=-1$) is the simplest candidate for such a component.
Although the concordance $\Lambda$CDM model is allowed by all current cosmological observations, it is also plagued by serious issues like fine-tuning and cosmic coincidence problems. Scalar field model \cite{scalar} is another example of dark energy where the equation of state $w$ evolves with time which in turn helps to solve the cosmic coincidence by tracker- type evolution. This also restricts the form of the potential for the scalar fields.

Another interesting approach to explain the late time acceleration of the universe is to modify the gravity at large scale ( infra-red modification of gravity). DGP brane-world model \cite{dgp} is one such example where the gravity is altered at large scales due to the slow leakage of gravitons from our observable universe ( modelled as an three brane) in to the higher dimensional bulk. The resulting Hubble equation can lead to late time acceleration in the observable universe. DGP model provides an interesting setup to modify gravity where one can construct a new four dimensional effective field theory  which contains nontrivial symmetry properties. These symmetries are due to the combination of five dimensional Poincare invariance and the brane paramatrization invariance. It has been shown that in this effective theory, there is a single scalar field $\pi$ which represents the position of the three-brane ( our observable universe) in the higher dimensional bulk. The effective action contains a cubic self-interaction term of the kind $(\partial \pi)^2 \Box\pi$ together with normal canonical kinetic energy term. This cubic self interaction term has the property that it leads to second order equation of motion of $\pi$. This term is also invariant under the {\it Galilean Transformation}:
\begin{eqnarray}
\pi &\rightarrow& \pi +a\nonumber\\
\partial_{\mu}\pi &\rightarrow& \partial_{\mu}\pi + b_{\mu} \label{piequation}
\end{eqnarray}

\noindent
where $a$ and $b_{\mu}$ are constants.  Due to the form of equation (\ref{piequation}) $\pi$ is often referred as Galileon field.
Later, a four dimensional theory for the field $\pi$ with Galileon symmetry was proposed that contains five terms which also had the inersting property that depsite the presence of higher derivatives in the action, the equations of motions are second order \cite{lutty, covariant} .  This causes the theory to be free of any ghosts. Cosmology with such an action has been widely studied in recent times \cite{galileon}. 

On the other hand, Dirac-Born-Infeld (DBI) action contains the lowest order dynamics of a brane embedded in a higher dimensional spacetime. This gives an interesting setup to study inflation \cite{dbiinf} and late time acceleration \cite{dbiquint,dbiessence} of the universe. 
If the universe is indeed described in a brane world scenario, then they should share the symmetries in the DBI action. Galileon terms can be thought of as a subset of all the higher derivative terms usually expected to be present in any effective field theory of the brane and they will be suppressed by some cut of scale (of different powers). In a recent work, 
Rham and Tolley \cite{rham} have constructed a general class of effective field theory which reproduces Galileon as well as the familiar DBI action under different limits. This has been extended by Goon et al. \cite{goon} where they have  constructed a general class of effective field theory assuming that a 3-brane is moving in the higher dimensional bulk. Galileon theory arises as a special case of this setup.  Similar generalization of Galileon field to maximally symmetric spacetime using a de-Sitter slicing has also been done in \cite{burrage}.

In this work, we study the late time acceleration of the universe in such a setup. We assume the simplest case where a maximally symmetric brane ( Minkowski Brane) is embedded in a maximally symmetric bulk (Minkowski bulk). We consider those three terms in the action which under small field limit generate the standard Galileon terms that arise in the DGP model under decoupling limit. We compare the cosmological behaviour in this setup to that of the concordance $\Lambda$CDM universe. We also constrain our model parameters using currently available observational data.

We start by giving a general introduction on DBI-Galileon model in section 2. In section 3, we discussed the late time cosmology in this model. We explore the constraints on the model parameters using observational data in section 4. Section 5 deals the conclusions.

\section{DBI-Galieleon}
\vspace{5mm}

We begin with a setup where the foliation is Gaussian normal with respect to the bulk metric $G_{AB}$ and the extrinsic curvature on each foliation is proportional to the induced metric.  This allows us to write the metric in the following form \cite{goon}:
\begin{equation}
G_{AB}dX^{A}dX^{B} = f(\pi)^{2}g_{\mu\nu}(x)dx^{\mu}dx^{\nu} + d\pi^{2}   \label{5Dlinelement}
\end{equation}

\noindent
where $\pi$ represents the Gaussian normal transverse coordinate, and $g_{\mu\nu}$ is the metric on the brane.
We have choosen a gauge used in \cite{goon}

\begin{equation}
X^{A}(x) |_{A=0...3} = x^{\mu} ; X^{4} = \pi(x).                                                  \label{gauge}
\end{equation}
Then the tangent vectors are given by
\begin{eqnarray}
e^{A}_{\mu} =\frac{\partial X^{A}}{\partial x^{\mu}} = \left \{ \begin{array}{ll}
\delta ^{\nu}_{\mu} & \mbox{$A = \nu$}  \\
\pi_{;\mu} & \mbox{$ A=4$} \end{array} \right.    \label{tangent}
\end{eqnarray}
\noindent
where, $\pi_{;\mu}$ denotes the covariant derivative of the field compatible to the metric $g_{\mu\nu}$.
The normal vectors are given by
\begin{eqnarray}
 n^{A} = \left\{ \begin{array}{ll}
 -\frac{1}{f^{2}} \gamma \pi^{;\mu} & \mbox{ $A = \mu$}  \\
\gamma & \mbox{$A = 4$} \end{array} \right.  \label{normal1}
\end{eqnarray}
where we have defined $\gamma$ as:
\begin{equation}
\gamma = \frac{1}{\sqrt{1+\frac{1}{f^{2}}(g^{\mu\nu}\pi_{;\mu}\pi_{;\nu})}}
\end{equation}
The induced metric on the $1+3$ dimensional hypersurface is defined as
\begin{equation}
h_{\mu\nu} = G_{AB}e^{A}_{\mu}e^{B}_{\nu}.
\end{equation}

\noindent
In the gauge given in equation(\ref{gauge}), the induced metric is

\begin{equation}
h_{\mu\nu} = f(\pi)^{2} g_{\mu\nu} + \pi_{;\mu}\pi_{;\nu}.                       \label{inducedmetric}
\end{equation}

\noindent

\noindent
Let us now consider that a Minkowski  brane ($M_{4}$) is embeddd in a Minkowski bulk ($M_{5}$). We choose cartesian co-ordinates ($x^{\mu}$, $\pi$) on $M_{5}$. Then, $\pi = constant$  will give the foliation of $M_{5}$ by $M_{4}$ and the five dimensional line element can be expressed as 

\begin{equation}
ds^{2} =  \eta_{\mu\nu}dx^{\mu}dx^{\nu} + d\pi^{2}
\end{equation} 
This is equivalent to putting
\begin{equation}
f(\pi) = 1, g_{\mu\nu} = \eta _{\mu\nu}.
\end{equation}
For this metric following Galileon terms are possible for which  Galileon symmetry is preserved and equation of motion for the field  $\pi$ is second order \cite{rham,goon}:

\begin{eqnarray}
& {\cal L}_{1} = &  \pi                                                                               \nonumber \\
& {\cal L}_{2} = &  - \sqrt{1 + (\partial \pi)^{2}}                                                     \nonumber \\
& {\cal L}_{3} = &  -[\Pi]^{2} + \gamma^{2}[\pi^{3}]                                                \nonumber \\
& {\cal L}_{4} = &  -\gamma ( [\Pi]^{2}-[\Pi^{2}])- 2\gamma^{3}([\pi^{4}]-[\Pi][\pi^{3}])                \label{dbiterms} \\
& {\cal L}_{5} = & -\gamma^{2}([\Pi]^{3}+2[\Pi^{3}] -3[\Pi][\Pi^{2}])-\gamma^{4}(6[\Pi][\pi^{4}]-6[\pi^{5}]-3([\Pi]^{2}-[|\Pi^{2}])[\pi^{3}]) \nonumber
\end{eqnarray}

\noindent
We have used the notation $\Pi$ for the matrix of second derivative of the field i.e. $\Pi_{\mu\nu} \equiv \pi_{;\mu;\nu}$. $[\Pi^{n}] $ has been used to denotes the trace of powers of the matrix. For example $[\Pi] = \pi_{;\mu}^{;\mu}$. $[\Pi^{2}] = \Pi_{\mu\nu}\Pi^{\mu\nu}$. All the indices has been raised with the metric $g^{\mu\nu}$. We have used the notation $[\pi^{n}]$ to denote the contraction of powers of $\Pi$ with $\pi_{;\mu}$. $[\pi^{n}] \equiv \pi_{;\mu}.\hspace{1mm}\Pi^{\mu(n-2)\nu}\hspace{1mm}.\pi_{;\nu}$. For example, $[\pi^{2}] = \pi_{;\mu}\pi^{;\mu}$, $[\pi^{3}] = \pi_{;\mu}\Pi^{\mu\nu}\pi_{;\nu}$. Here also all the indices has been raised with the metric $g^{\mu\nu}$.\\

These terms in equation (\ref{dbiterms}) are the  so called DBI Galileon terms \cite{rham,goon}.  Applying the small field limit to these DBI Galileon terms, one gets back the original Galileon terms \cite{rham,goon}:

\begin{eqnarray}
&{\cal L}_{1} = & \pi \nonumber\\
&{\cal L}_{2} = & -\frac{1}{2}(\partial \pi)^2 \nonumber\\
&{\cal L}_{3} = & -\frac{1}{2}(\partial \pi)^2 \left[ \Pi\right] \nonumber\\
&{\cal L}_{4} = & -\frac{1}{2}(\partial \pi)^2 \left(\left[ \Pi\right]^2 - \left[\Pi ^2\right] \right)\nonumber\\
&{\cal L}_{5} = & -\frac{1}{2}(\partial \pi)^2 \left(\left[ \Pi\right]^3 - 3\left[\Pi\right]\left[\Pi ^2\right] + 2\left[\Pi^3\right]\right) 
\end{eqnarray}

The first three terms in the standard Galileon case, are the simplest ones representing a potential term, a canonical kinetic term, and a self interaction term that can also arise in the decoupling limit of DGP model. Cosmology in FRW background universe with these terms has been studied by various author \cite{galileon}. 

\section{Late time Cosmological Evolution with DBI-Galileon}
\vspace{5mm}

In our present study, we consider the corresponding three terms in the DBI Galileon case (${\cal L}_{1}, {\cal L}_{2}$ and ${\cal L}_{3}$ in eqn (\ref{dbiterms})) and study the cosmology in a FRW background.  We also replace the $\pi$ term in the ${\cal L}_{1}$ by a more general potential function $V(\pi)$. The corresponding Einstein's equations are given by

\begin{equation}
3H^{2} = \rho_{m} + V(\pi) + \frac{c_{2}}{\sqrt{1-\dot{\pi}^{2}}}                               \label{einstein00}
\end{equation}
and,
\begin{equation}
2\dot{H} + 3H^{2} = V(\pi)+c_{2}\sqrt{1-\dot{\pi}^{2}}-\frac{c_{3}\dot{\pi}^{2}\ddot{\pi}}{1-\dot{\pi}^{2}}   \label{einsteinii}
\end{equation}

We have set $8\pi G = 1$. Equation of motion of the field $\pi$ is given by
\begin{equation}
c_{2}\frac{\ddot{\pi}}{(1-\dot{\pi}^{2})^{3/2}}+ c_{2}\frac{3H\dot{\pi}}{\sqrt{1-\dot{\pi}}} + c_{3}\frac{3H\dot{\pi}\ddot{\pi}}{(1-\dot{\pi}^{2})} + V(\pi)_{,\pi}=0                                                                  \label{Eompi}
\end{equation}
where $V(\pi)_{,\pi}$ denotes the derivative of the potential with respect to the field. $c_{2}$ and $c_{3}$ are the corresponding coupling constants for ${\cal L}_{2}$ and ${\cal L}_{3}$.

We define the following dimensionless variables:

\begin{equation}
x =  \dot{\pi},  \hspace{2mm} y =  \frac{\sqrt{V}}{\sqrt{3}H}, \hspace{2mm} \lambda =  -\frac{1}{H_{0}}\frac{V^{'}}{V},
\end{equation}
\begin{equation}
h = \frac{H}{H_{0}} , \hspace{2mm} \Gamma =  \frac{VV_{,\pi\pi}}{(V_{,\pi})^{2}} ,
\end{equation}
\begin{equation}
\alpha =  \frac{c_{2}}{H_{0}^{2}} ,  \hspace{2mm} \beta =  \frac{c_{3}}{H_{0}}.     
\end{equation}

\noindent
Here $H_{0}$ represents the Hubble constant at present. In terms of  these dimensionless variables, we form the following autonomous system of equation:

\begin{equation}
x^{'} = \frac{3\lambda y^{2}h(1-x^{2})^{3/2}-3\alpha x(1-x^{2})}{\alpha + 3\beta h x \sqrt{1-x^{2}}}   \label{xautonomus}
\end{equation}

\begin{equation}
y^{'} = \frac{3y}{2}\left[1-\frac{\lambda x}{3h} -y^{2}-\frac{\alpha\sqrt{1-x^{2}}}{3h^{2}}-\frac{\beta x^{2}}{3h(1-x^{2})}\left\{\frac{3\lambda y^{2}h(1-x^{2})^{3/2}-3\alpha x(1-x^{2})}{\alpha + 3\beta h x \sqrt{1-x^{2}}}\right\}\right]                                      \label{yautonomus}
\end{equation}                                                                                

\begin{equation}
h^{'} = \frac{3h}{2}\left[y^{2}-1+\frac{\alpha\sqrt{1-x^{2}}}{3h^{2}}+\frac{\beta x^{2}}{3h(1-x^{2})}\left\{\frac{3\lambda y^{2}h(1-x^{2})^{3/2}-3\alpha x(1-x^{2})}{\alpha + 3\beta h x \sqrt{1-x^{2}}}\right\}\right]                                      \label{hautonomus}
\end{equation}
\begin{equation}
\lambda ^{'} = -\frac{\lambda^{2}x}{h}(\Gamma -1)                                   \label{lambdaautonomus}
\end{equation}

Here, prime denotes derivative w.r.t $ln(a)$. The equation of state parameter $w_{\pi}$ for the field $\pi$ is given by
\begin{equation}
w_{\pi} = \frac{-1}{\sqrt{1-x^{2}}}\left\{\frac{3h^{2}y^{2}(1-x^{2}) + \alpha(1-x^{2})^{3/2}-\beta x^{2}h x^{'}}{\alpha + 3h^{2}y^{2}\sqrt{1-x^{2}}}\right\}
\end{equation}

We evolve the system from the decoupling era ($a \approx 10^{-3}$) to the present day ($ a = 1$). In the beginning we assume that there was negligible contribution from the $\pi$ field and the universe was dominated only by matter. We also assume that the field $\pi$ was initially frozen due to large Hubble friction. This is similar to the thawing class of models previously studied for both scalar fields \cite{scherrer, soma} and galileon fields \cite{wali}. This sets the initial condtions for $h$ and $x$ as
\begin{equation}
h_{i} = 10^4 \sqrt{10 \Omega_{m0}}, \hspace{2mm} x_{i} \approx 0.
\end{equation}

\begin{figure}
\begin{center}
\begin{tabular}{|c|c|}
\hline
{\includegraphics[width=2.6in,height=2in,angle=0]{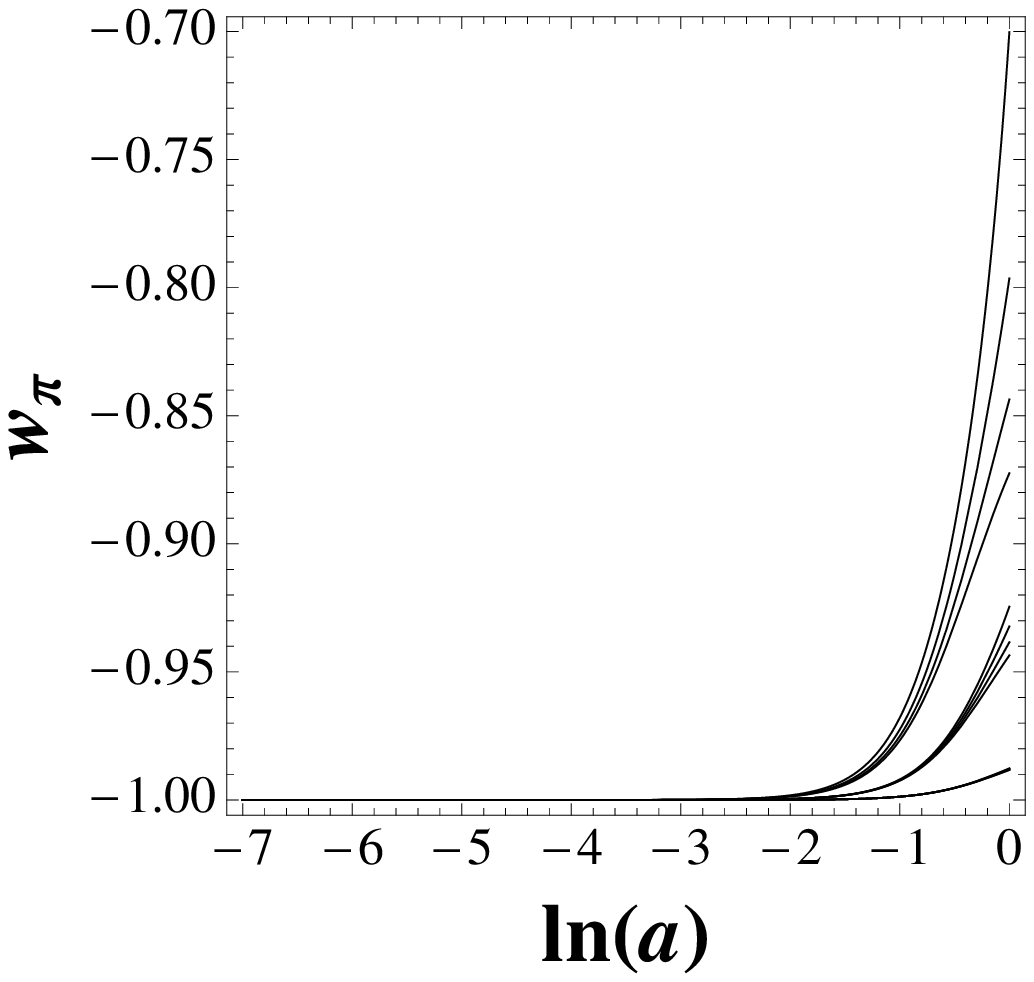}}&
{\includegraphics[width=2.6in,height=2in,angle=0]{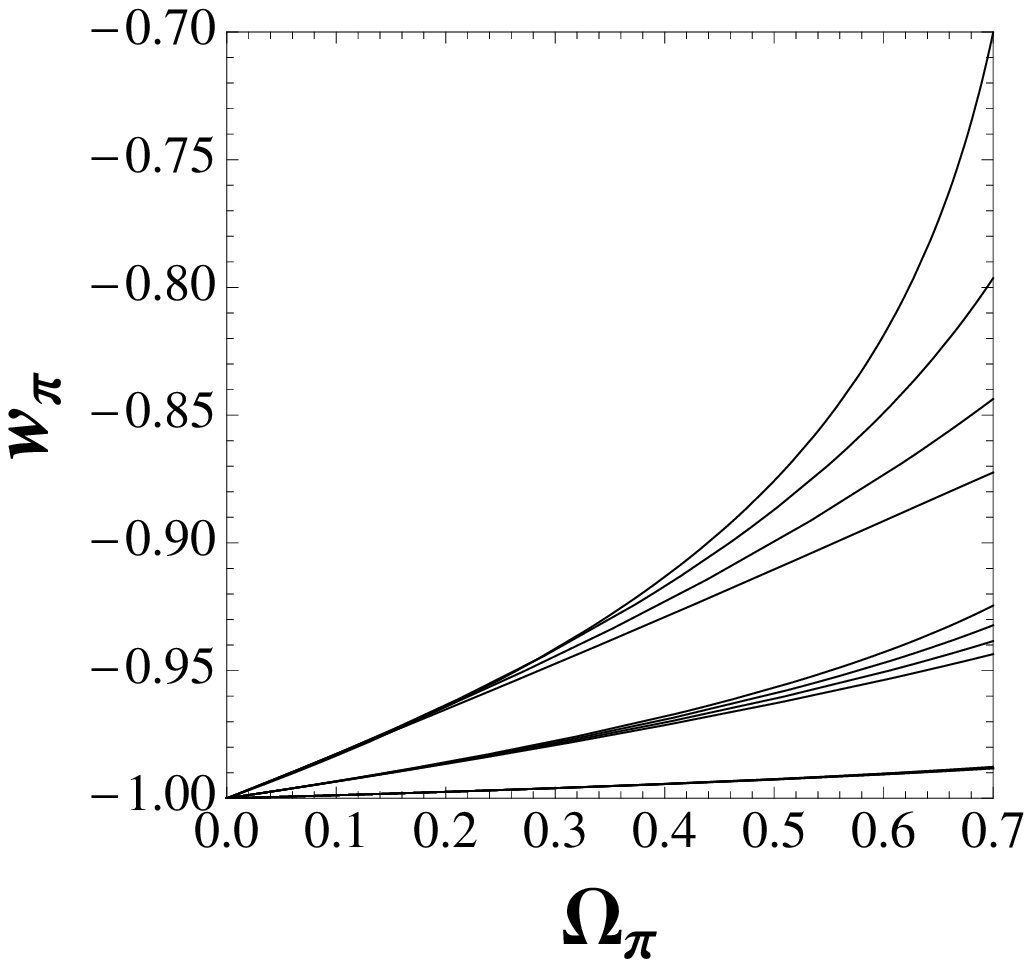}}\\
\hline
 \end{tabular}
\caption{Plot of $w_{\pi}$vs ln(a) (left)  and  $\Omega_{\pi}$ (right). In each plot, the three sets are for $\lambda_{i} = 0.2, 0.5, 0.9$ from bottom to top. In each set,  $V(\pi) = \pi, \pi^2, e^{\pi}$ and $1/\pi^2$ from top to bottom. $\Omega_{mo} =0.3$ and $\alpha=0.3$, $\beta = 0.1$.}
\end{center}
\end{figure}
\noindent 
The initial condtion for $y$ is set in the following way: using eqn (\ref{einstein00}), one can write
\begin{equation}
y_{i} ^{2} = \Omega_{\pi i} - \frac{\alpha}{3 h_{i}^2 \sqrt{1-x_{i}^2}}.
\end{equation}

This allows us to replace $y_{i}$ in terms of other quantities. On the other hand, $\Omega_{\pi i}$ is related to $\Omega_{\pi 0}$ which in turn is equal to  $1-\Omega_{m0}$ due to the flatness condition. Hence $y_{i}$ is not an independent quantity but is related to other model parameters like $\alpha$, and $\Omega_{m0}$. The initial value for $\lambda$, $\lambda_{i}$, is an interesting parameter. It controls the deviation from the $\Lambda$CDM behaviour.  This is shown in Figure 1. It is evident from this figure that for smaller values of $\lambda_{i}$ models with different potentials can hardly be  distinguished from each other as well as from $w=-1$. As one increases $\lambda_{i}$, they start deviating from $w=-1$ and deviate from each other as well. Moreover the equation of state $w_{\pi}$, for linear potential has the largest deviation from $w=-1$. Note that this potential arises naturally in the DBI Galileon model. This result is also similar to the standard scalar field \cite{soma} as well standard Galileon field models \cite{wali}.   

One can also Taylor expand the scale factor of the universe around the present era ($t=t_{0}$) as follows:
\begin{equation}
a(t) = a(t_{0}) + a(t_{0})\sum^{\infty}_{n=1} \frac{\gamma_{n}(t_{0})}{n!}\left[H_{0}(t-t_{0})\right]^n,
\label{state}
\end{equation} 
where,
\begin{equation}
\gamma_{n} = \frac{d^{n}a}{dt^n}/(aH^{n}).
\label{gamma}
\end{equation}
 It is now straightforward to show that $-\gamma_{2} \equiv q$  is the deceleration parameter. Similarly $\gamma_{3}$ is related to the Statefinder $r$ or jerk $j$ and $\gamma_{4}$ is related to the snap $s$ and so on. Using the equations (\ref{state}) and (\ref{gamma}), one can now construct the Statefinder Hierarchy $S_{n}$ \cite{varun} as:
 \begin{eqnarray}
 S_{2} &=& \gamma_{2} + \frac{3}{2}\Omega_{m}\\
 S_{3} &=& \gamma_{3}\\
 S_{4} &=& \gamma_{4} + \frac{3^2}{2}\Omega_{m}\\
 S_{5} &=& \gamma_{5} - 3 \Omega_{m} - \frac{3^3}{2}\Omega^{2}_{m}\\
 S_{6} &=& \gamma_{6} + \frac{3^3}{2}\Omega_{m} + {3^4}\Omega^{2}_{m} + \frac{3^4}{4}\Omega^{3}_{m}  \hspace{2mm} \hspace{2mm} and \hspace{2mm} so \hspace{2mm} on.
 \end{eqnarray}

\begin{figure}
\begin{center}
\begin{tabular}{|c|c|}
\hline
 & \\
{\includegraphics[width=2.6in,height=2in,angle=0]{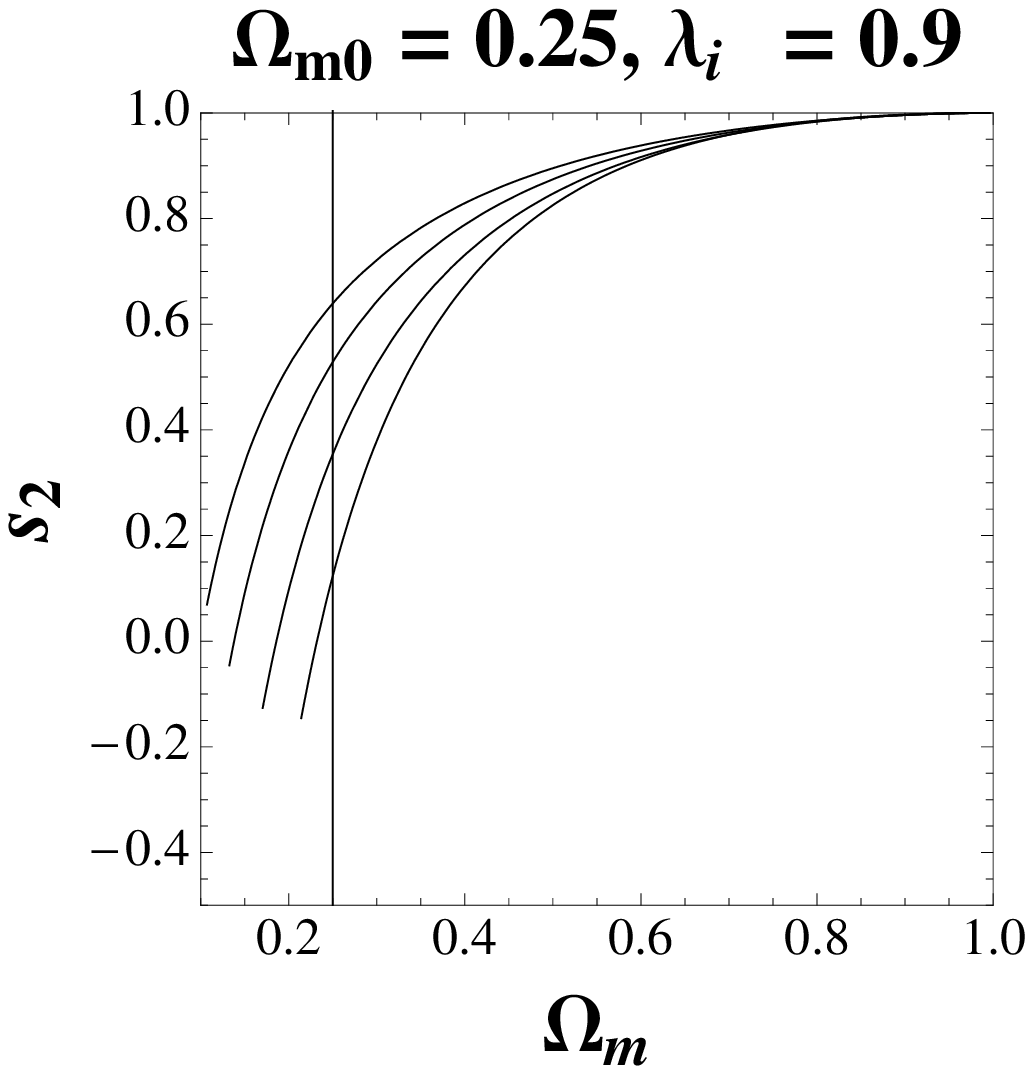}}&
{\includegraphics[width=2.6in,height=2in,angle=0]{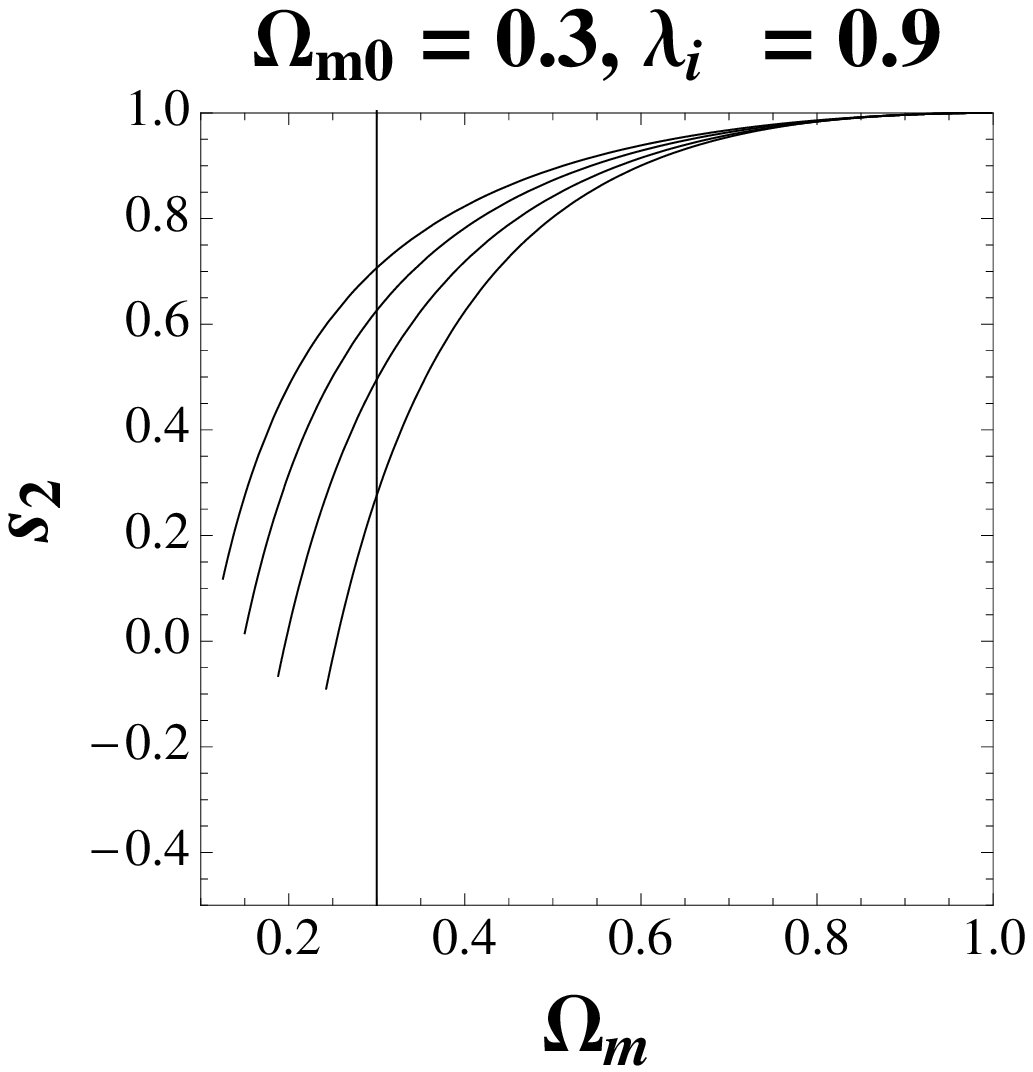}}
\\
\hline
{\includegraphics[width=2.6in,height=2in,angle=0]{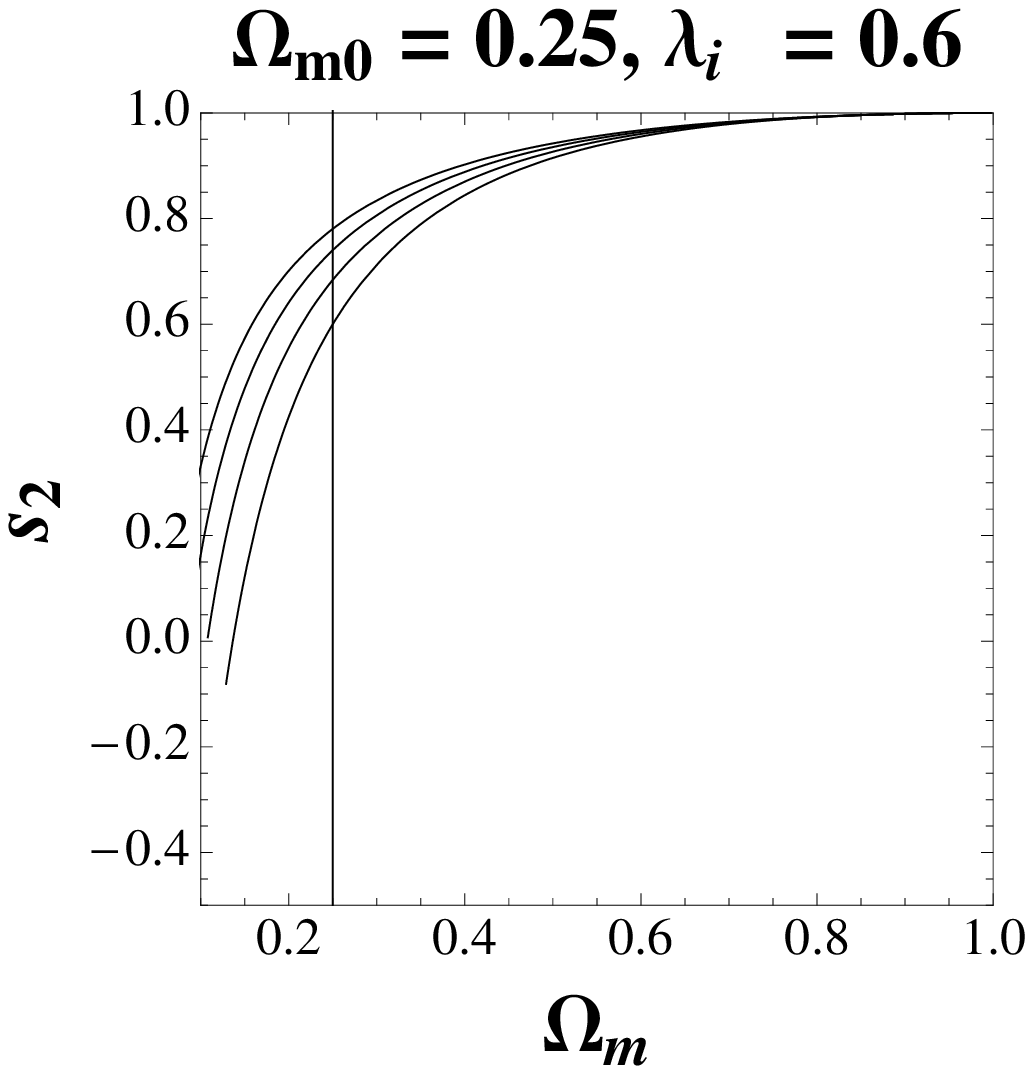}}&
{\includegraphics[width=2.6in,height=2in,angle=0]{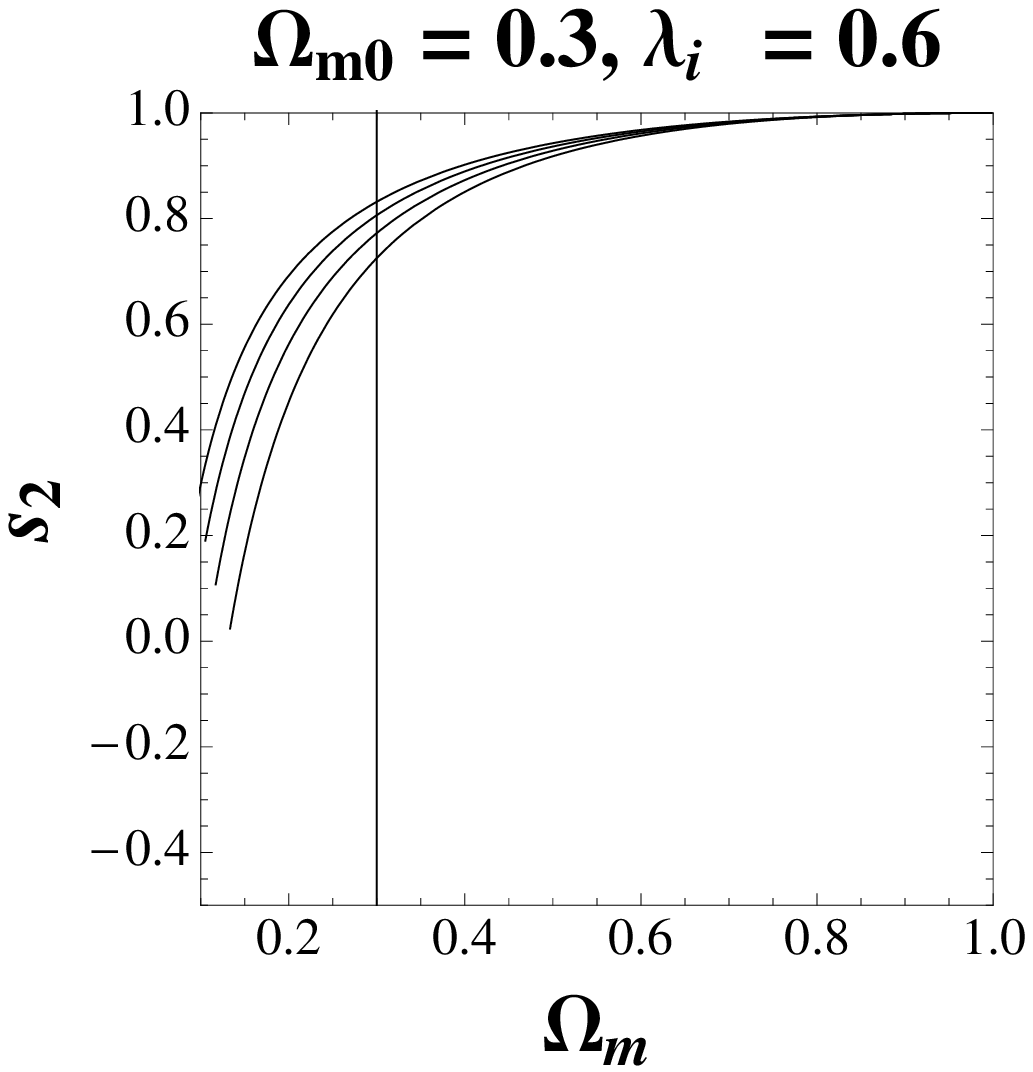}}
\\
\hline
\end{tabular}
\caption{Phase space diagram in $S_{2}-\Omega_{m}$ plane for different potentials. $V(\pi) = \pi, \pi^2, e^{\pi}$ and $\frac{1}{\pi^2}$ from bottom to top. $\alpha=0.3$ and $\beta=0.3$ for all the plots. The vertical line represents the present day ($z=0$).}
\end{center}
\end{figure}

\noindent
It is interesting to note that for $\Lambda$CDM, $S_{n} = 1$ throughout the history of the Universe.  Hence any deviation from $S_{n} =1$ represents models with non-$\Lambda$CDM behaviour. $S_{n}$'s can also be used to study degeneracies between different dark energy models. We now use these Statefinder Hierarchies to study DBI-Galileon models with differenet potentials.

In Figure (2), we show the behaviour in the $S_{2}-\Omega_{m}$ plane for different potentials and for different values of $\lambda_{i}$. As in the case of the equation of state, here too the deviation from the $\Lambda$CDM is small for smaller $\lambda_{i}$, and large for bigger $\lambda_{i}$. Also the deviation from the $\Lambda$CDM behaviour is highest for the linear potential. The degeneracies between models with different potentials is higher for smaller $\lambda_{i}$. Also for smaller values of $\Omega_{m0}$, deviations from the $\Lambda$CDM behaviour as well as deviations between different potentials are higher. In this figure, we keep both $\alpha$ and $\beta$ fixed at $0.3$. For other values, one can get similar behaviour. The dependence on $\alpha$ and $\beta$ is discussed below.

\begin{figure}
\begin{center}
\begin{tabular}{|c|c|}
\hline
 & \\
{\includegraphics[width=2.6in,height=2in,angle=0]{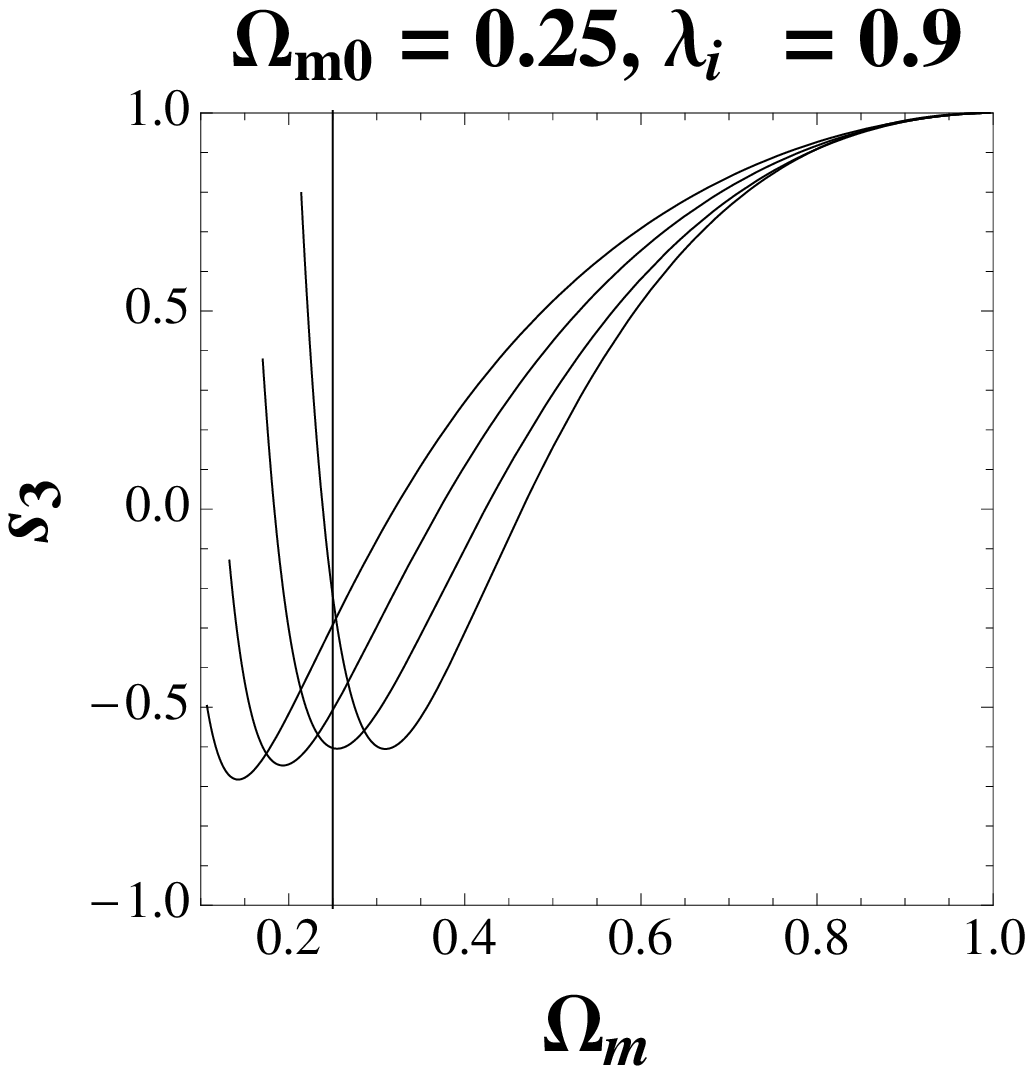}}&
{\includegraphics[width=2.6in,height=2in,angle=0]{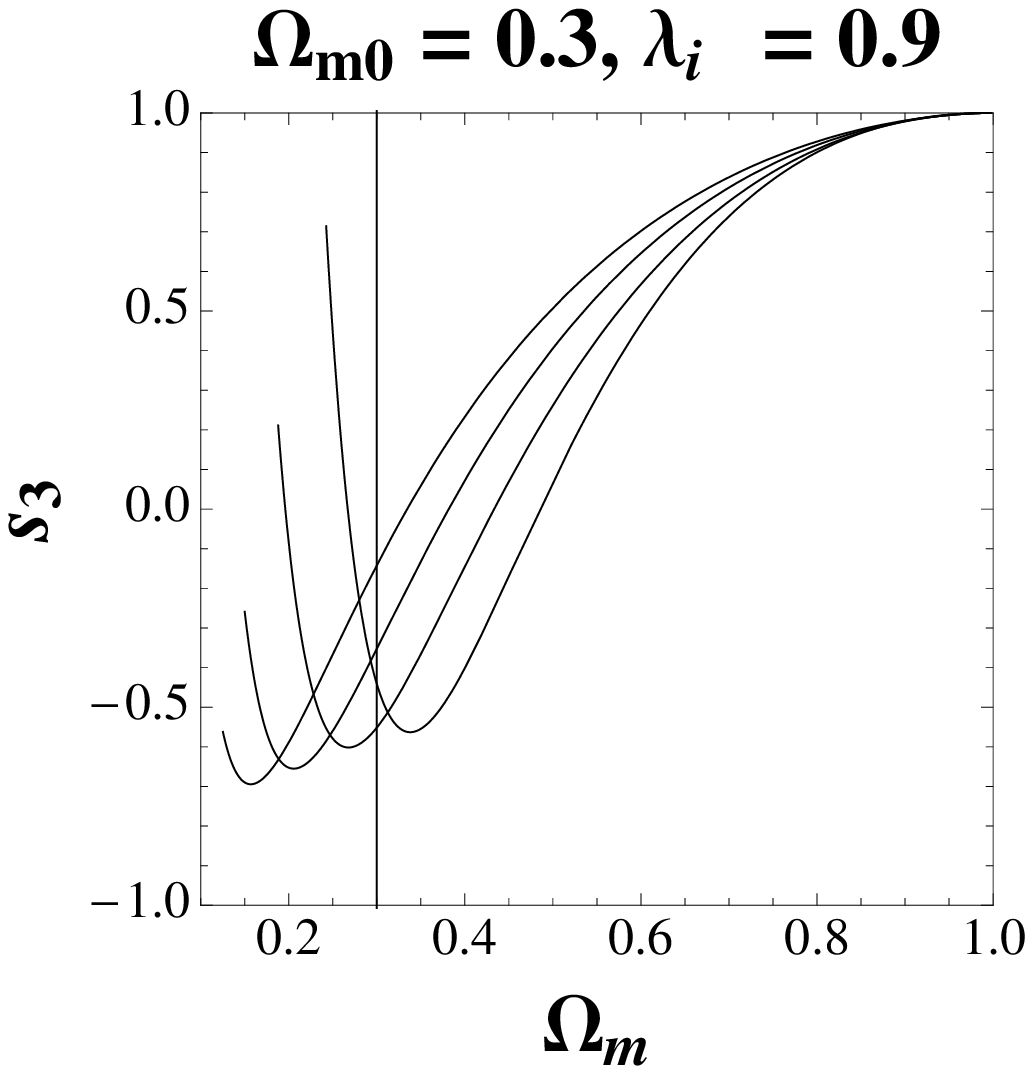}}
\\
\hline
{\includegraphics[width=2.6in,height=2in,angle=0]{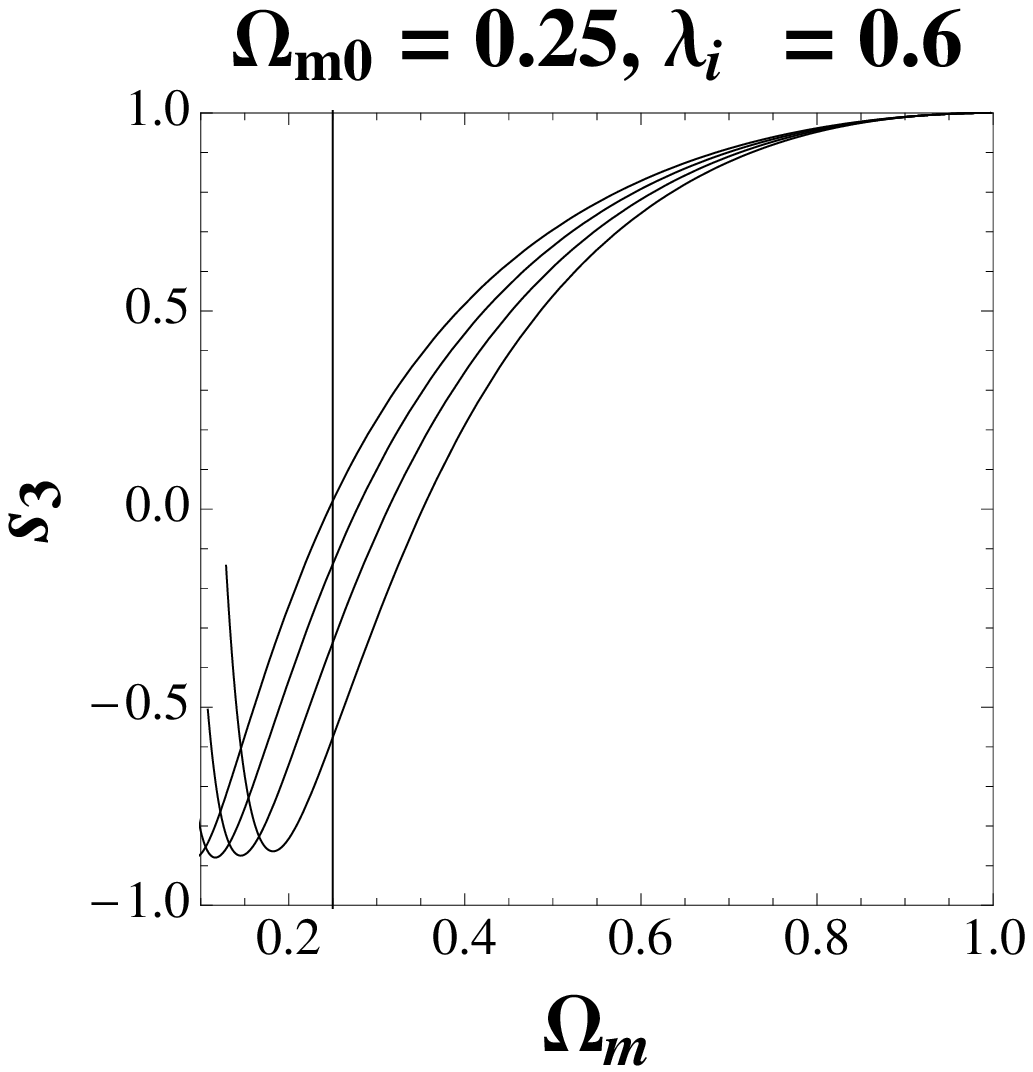}}&
{\includegraphics[width=2.6in,height=2in,angle=0]{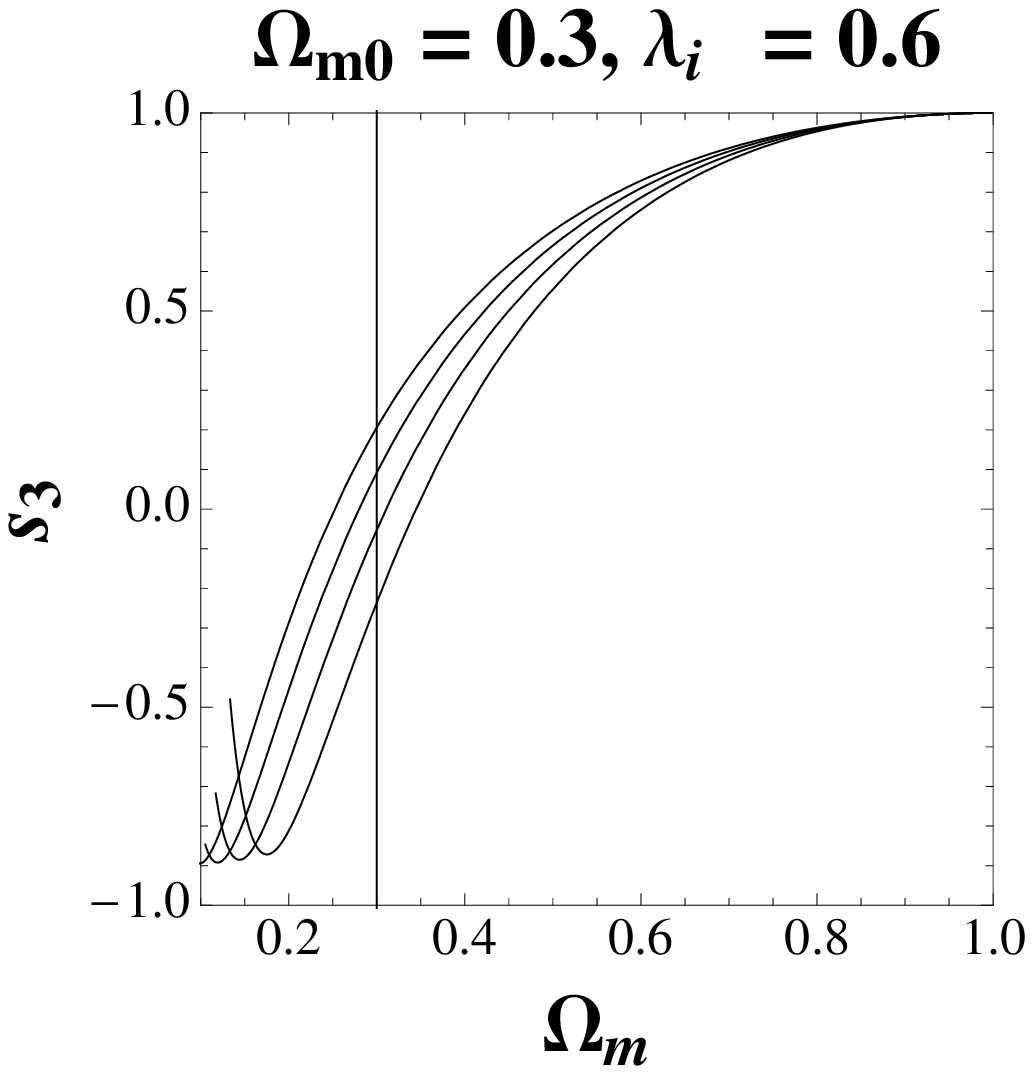}}
\\
\hline
\end{tabular}
\caption{Phase space diagram in $S_{3}-\Omega_{m}$ plane for different potentials. $V(\pi) = \pi, \pi^2, e^{\pi}$ and $\frac{1}{\pi^2}$ from bottom to top. $\alpha=0.3$ and $\beta=0.3$ for all the plots. The vertical line represents the present day ($z=0$).}
\end{center}
\end{figure}

Next we study the evolution for different potentials in the $S_{3}-\Omega_{m}$ phase-space in Figure (3). For this, we also fix $\alpha = 0.3$ and $\beta=0.3$.  This figure is interesting. Here for $\lambda_{i}$ being $0.9$ and $0.6$, different potentials deviates more from each other at present($z=0$) for $\lambda_{i} = 0.6$.  This is opposite to what happens in the case of $S_{2}-\Omega_{m}$ phase space where the deviation at present is higher for $\lambda_{i} = 0.9$. We have also demonstrated that all the potentials converge to $\Lambda$CDM for very small values of $\lambda_{i}$ ($\lambda_{i} \sim 0.1$) for which all $S_{n}$'s are exactly $1$.  This shows that  there exists some $\lambda_{i}$ which optimizes the deviation in $S_{3}$ for different potentials.

In Figure (4), we show that phase space is $S_{3}-S_{2}$ plane. Here also the models with different potentials deviate more for smaller $\Omega_{m0}$ and higher $\lambda_{i}$. the deviation between models in this phase space  is also higher than the two phase spaces.

\begin{figure}
\begin{center}
\begin{tabular}{|c|c|}
\hline
 & \\
{\includegraphics[width=2.6in,height=2in,angle=0]{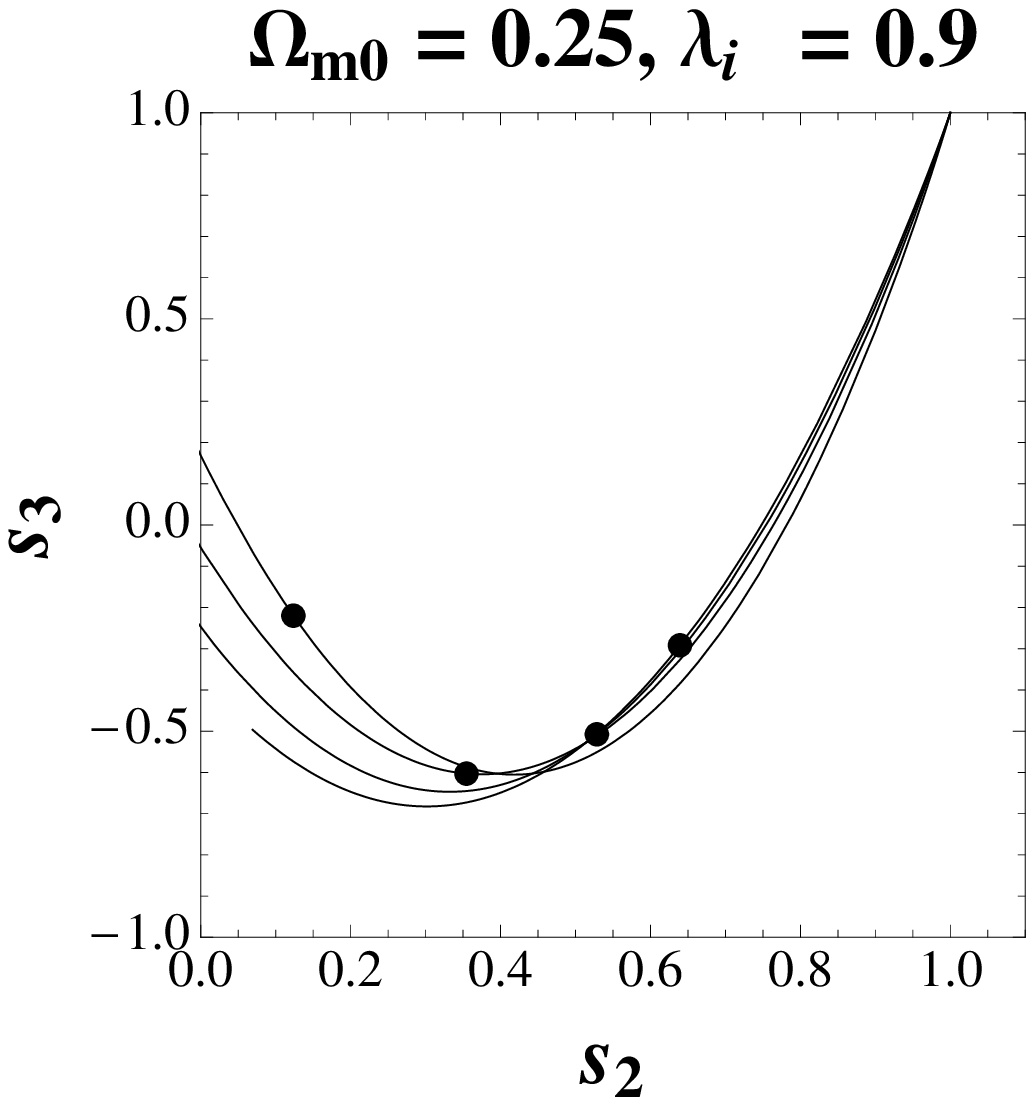}}&
{\includegraphics[width=2.6in,height=2in,angle=0]{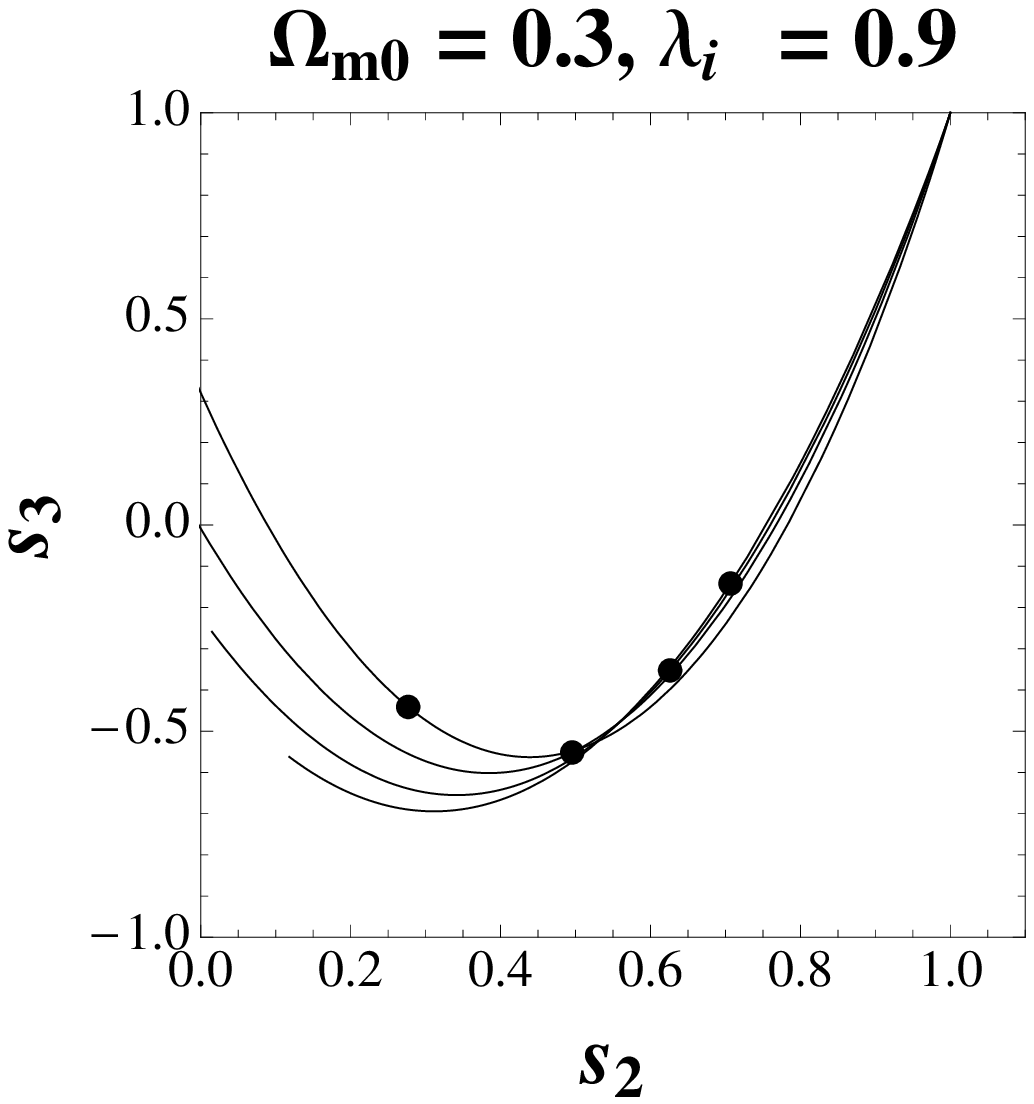}}
\\
\hline
{\includegraphics[width=2.6in,height=2in,angle=0]{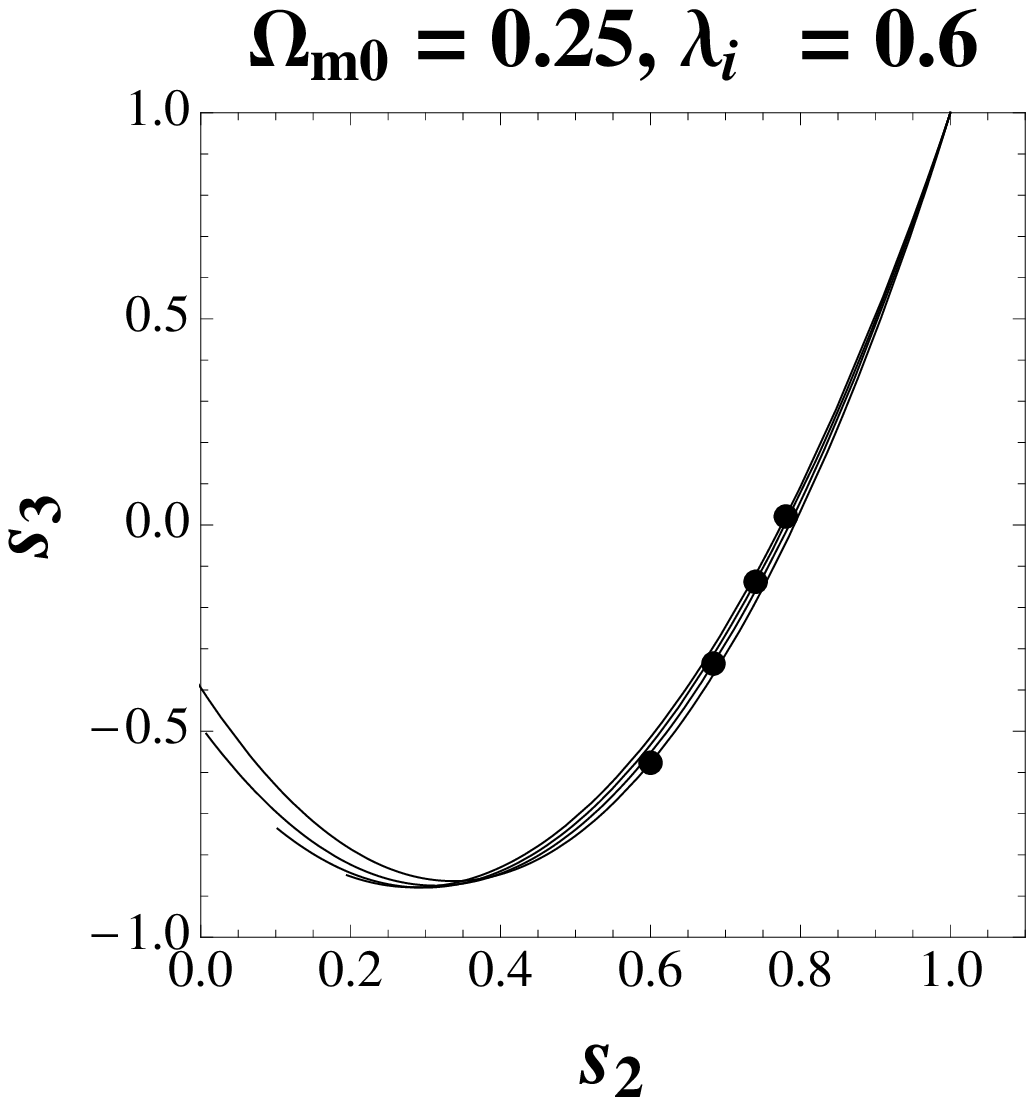}}&
{\includegraphics[width=2.6in,height=2in,angle=0]{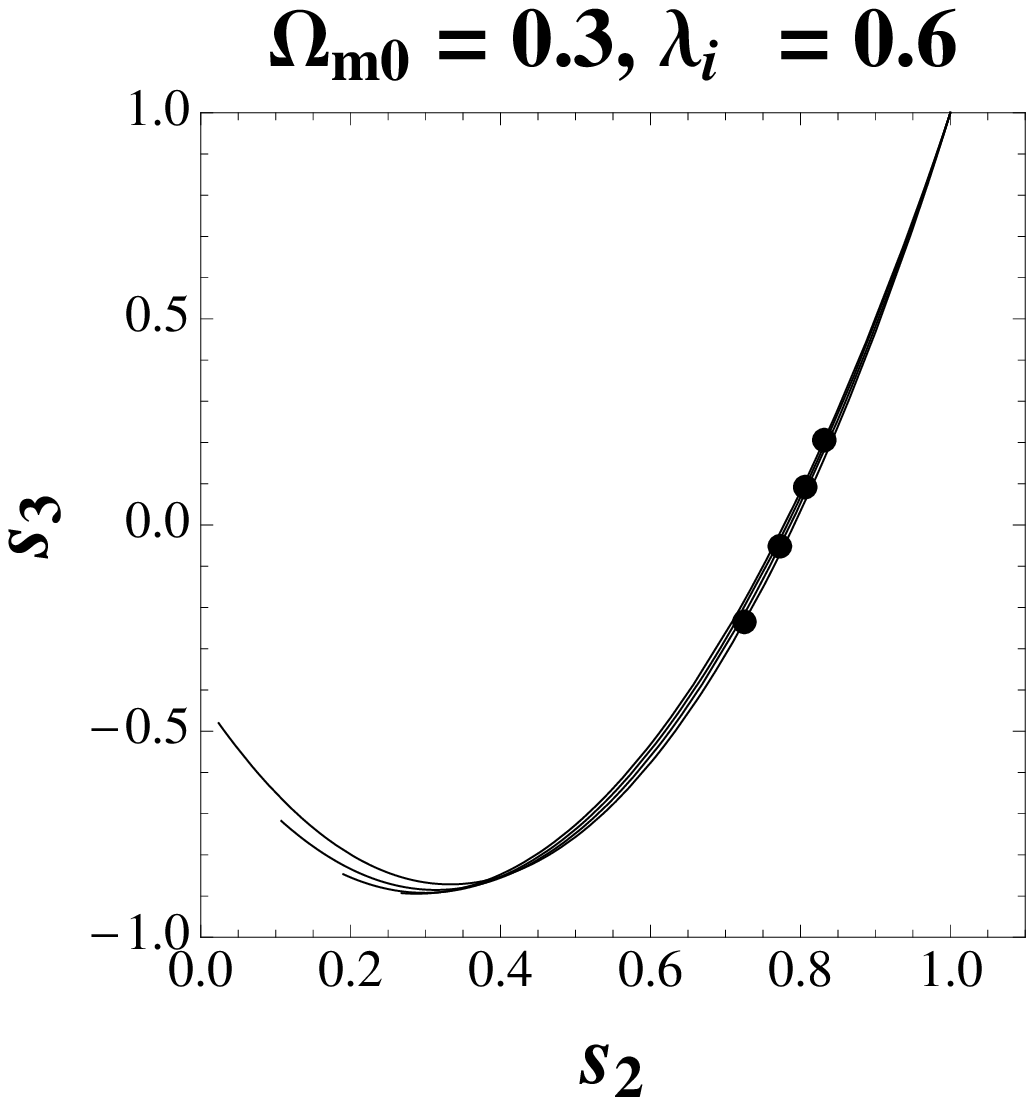}}
\\
\hline
\end{tabular}
\caption{Phase space diagram in $S_{3}-S_{2}$ plane for different potentials. $V(\pi) = \pi, \pi^2, e^{\pi}$ and $\frac{1}{\pi^2}$ from top to bottom at the left side of each figure. $\alpha=0.3$ and $\beta=0.3$ for all the plots.,The dots represents the present day ($z=0$).}
\end{center}
\end{figure}

In Figure (5), we show the dependence on the two parameters $\alpha$ and $\beta$.  In this case we fix $\lambda_{i} =0.9$ and $\Omega_{m0} = 0.3$. This is because, this choice maximised the deviations between different potentials as shown in Figure (4). Figure (5) shows that deviations between different potentials are larger for smaller values of both $\alpha$ and $\beta$. For some choices of $\alpha$ and $\beta$, there is complete degeneracy between some of the potentials. This is evident for the case $\alpha=0.3$ and $ \beta=0.8$ where the linear and the quadratic potentials are fully degeberate. 

From all these phase space diagrams, one can conclude that different potentials deviate maximum from each other for smaller $\Omega_{m0}$ and larger $\lambda_{i}$. But this can be adjusted by suitably choosing $\alpha$ and $\beta$ to make them completely degenerate.

\begin{figure}
\begin{center}
\begin{tabular}{|c|c|}
\hline
 & \\
{\includegraphics[width=2.6in,height=2in,angle=0]{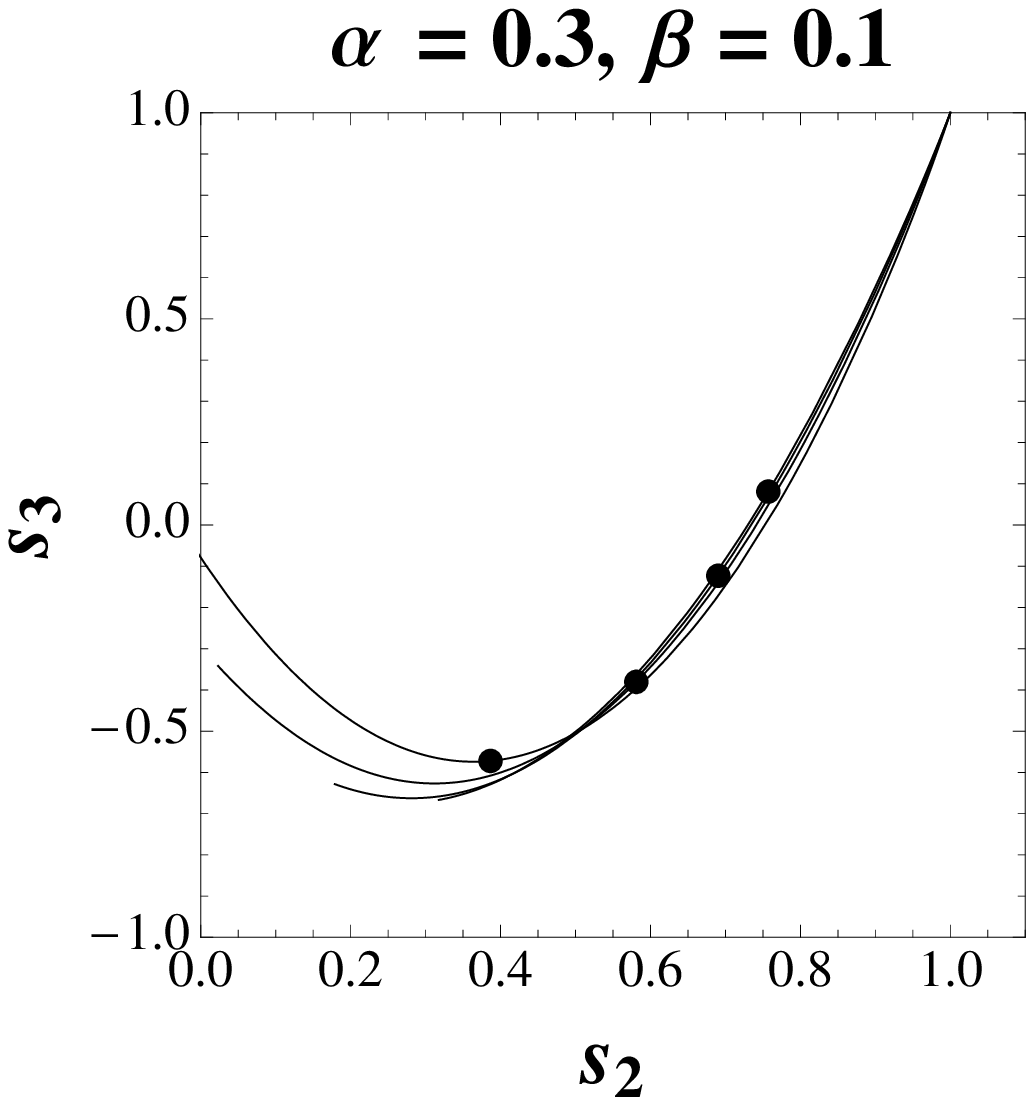}}&
{\includegraphics[width=2.6in,height=2in,angle=0]{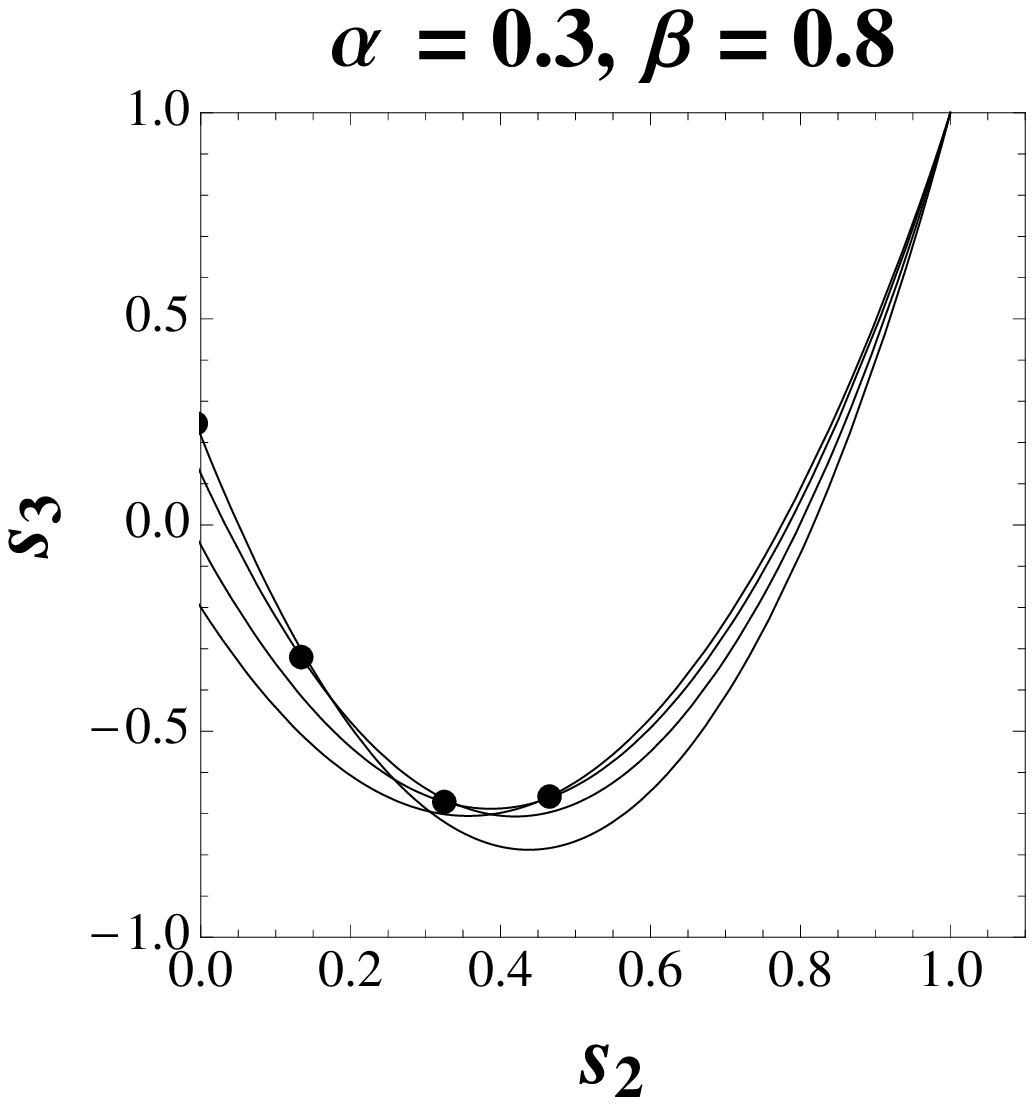}}
\\
\hline
{\includegraphics[width=2.6in,height=2in,angle=0]{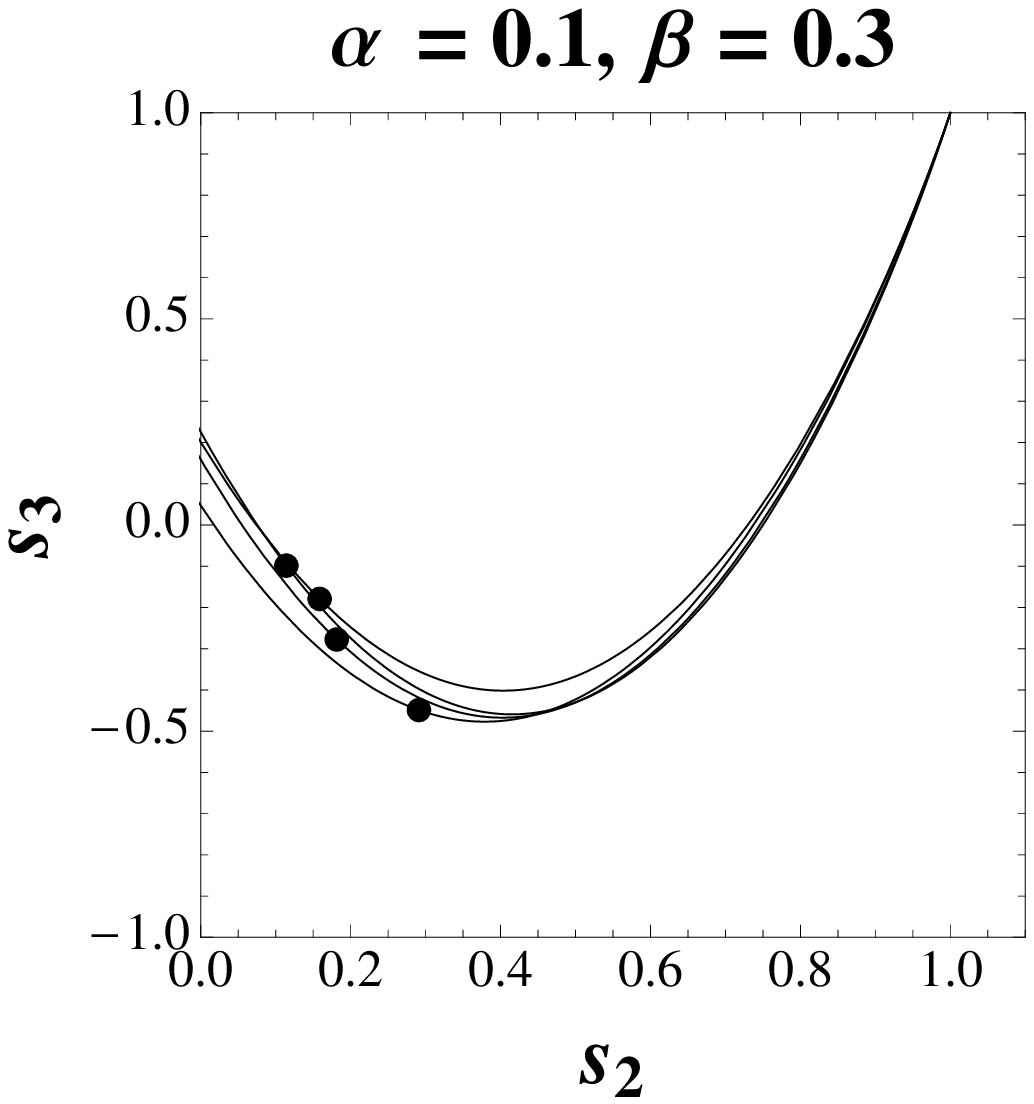}}&
{\includegraphics[width=2.6in,height=2in,angle=0]{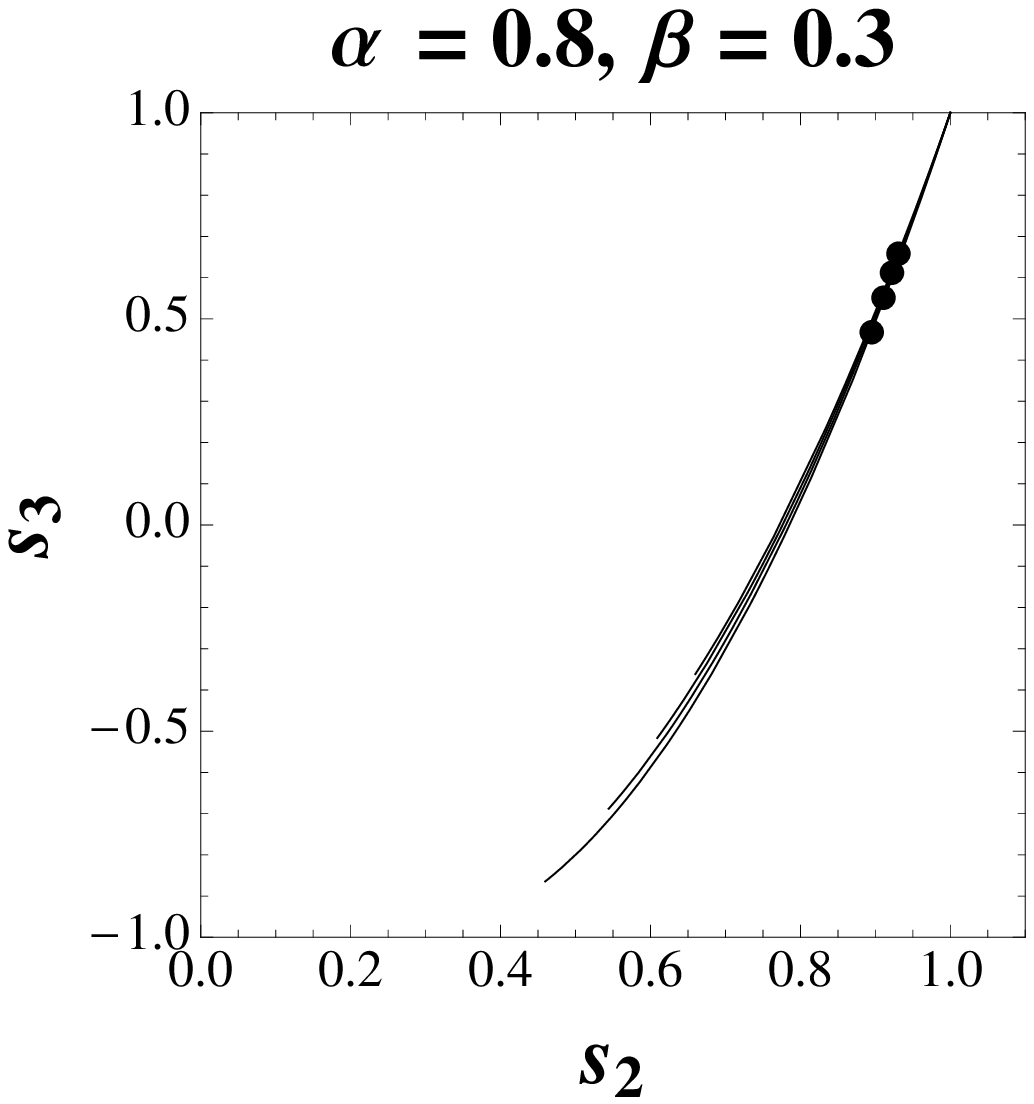}}
\\
\hline
\end{tabular}
\caption{Phase space diagram in $S_{3}-S_{2}$ plane for different potentials. $V(\pi) = \pi, \pi^2, e^{\pi}$ and $\frac{1}{\pi^2}$ from top to bottom at the left side of each figure. $\lambda_{i}=0.9$ and $\Omega_{m0}=0.25$ for all the plots. The dots represents the present day ($z=0$).}
\end{center}
\end{figure}

\section{Observational Constraints}

With this in mind, we now put constraints on these four parameters ($\lambda_{i}$, $\Omega_{m0}$, $\alpha$ and $\beta$)  by using recent observational data.

To start with, we consider the Type Ia supernova observation which is one of the direct probes for the cosmological expansion. In this case, one measures the apparent luminosity of the supernova explosion from the photon flux received. Cosmologically the relevant quantity is the luminosity distance $d_{L}(z)$ defined  as: 

\begin{equation}
d_{L}(z) = (1+z)\int_0^z\frac{dz^{\prime}}{H(z^{\prime})}.
\end{equation}

The distance modulus $\mu$ (which is an observable quantity) is related to the luminosity distance as
\begin{equation}
\mu = m-M = 5\log\frac{d_{L}}{Mpc}+25,
\end{equation}

\noindent
where m and M are the apparent and absolute magnitudes of the Supernovae respectively. We consider the latest Union2.1 data compilation \cite{suzuki} consisting of 580 data points for the observable $\mu$.

Next, we use the observational data on Hubble parameter as recently compiled by Moresco et al. \cite{Moresco2012} in the redshift range $0 < z < 1.75$. The sample contains 19 observational data point for $H(z)$ spanning almost $10$ Gyr of cosmic evolution. These values are given in Table 1. It also contains the  latest measurement of the Hubble constant $H_{0}$ \cite{komat}.

\begin{table}[h!]
\begin{center}
\begin{tabular}{llll}
\hline \hline
$z$ & $H(z)$ & $\sigma_{H(z)}$ & Ref.\\
\hline
0.090 & 69 & 12 & \cite{Simon2005}\\
0.170 & 83 & 8 & \cite{Simon2005}\\
0.179 & 75 & 4 & \cite{Moresco2012}\\
0.199 & 75 & 5 & \cite{Moresco2012}\\
0.270 & 77 & 14 & \cite{Simon2005}\\
0.352 & 83 & 14 & \cite{Moresco2012}\\
0.400 & 95 & 17 & \cite{Simon2005}\\
0.480 & 97 & 62 & \cite{Stern2010}\\
0.593 & 104 & 13 & \cite{Moresco2012}\\
0.680 & 92 & 8 & \cite{Moresco2012}\\
0.781 & 105 & 12 & \cite{Moresco2012}\\
0.875 & 125 & 17 & \cite{Moresco2012}\\
0.880 & 90 & 40 & \cite{Stern2010}\\
1.037 & 154 & 20 & \cite{Moresco2012}\\
1.300 & 168 & 17 & \cite{Simon2005}\\
1.430 & 177 & 18 & \cite{Simon2005}\\
1.530 & 140 & 14 & \cite{Simon2005}\\
1.750 & 202 & 40 & \cite{Simon2005}\\
\hline \hline
\end{tabular}
\caption{$H(z)$ measurements (in units [$\mathrm{km\,s^{-1}Mpc^{-1}}$]) and their errors.}
\label{tab:HzBC03}
\end{center}
\end{table}

Lastly, we consider the combined BAO/CMB constraints as  recently derived by Giostri et al. \cite{giostri12}.

\noindent
We start defining the comoving sound horizon at the decoupling as:

\begin{equation}
r_s(z_*)= \frac{H_{0}}{\sqrt{3}} \int_0^{1/(1+z_*)}\frac{da}{a^2h(a)\sqrt{1+(3\Omega_{b0} / 4 \Omega_{\gamma 0})a} },\label{eq:r_s}
\end{equation}
where $\Omega_{\gamma 0}$ and $\Omega_{b0}$ are the photon and baryon density parameter at preset respectively. We have assumed the speed of light $c=1$ in our calculations. $z_*$ is the redshift at decoupling and is given by the formula derived by Hu and Sugiyama\cite{hu96}. Taking the data from WMAP7 \cite{jarosik11}, we put $z_* = 1091$ exactly. We also fix the redshift of the drag epoch at $z_d \approx 1020$.

We next define the acoustic scale:

\begin{equation}
l_A=\pi\frac{d_A(z_*)}{r_s(z_*)}\quad,
\end{equation}
where  $d_A(z_*)= H_{0}\int_{0}^{z_*}dz'/h(z')$ is the comoving angular-diameter distance. The dilation scale is also defind as \cite{eisen}:

\begin{equation}  
D_V(z):=\left[ d_A^2(z) H_{0}z/h(z) \right]^{1/3}
\end{equation}

The 6dF Galaxy Survey \cite{beutler} and the WiggleZ team \cite{blake11} have measured this quantity  at $0.106$, $z=0.44$, $z=0.60$ and $z=0.73$. Percival et al.  \cite{Percival} also measured $\frac{r_{s}(z_{d})}{D_{V}(z)}$ at $z=0.2$ and $z=0.35$. One can combine these masurements with the WMAP-7 measurment of $\l_{a}$ \cite{jarosik11} and can obtain the combined meaurement of BAO/CMB for the quantity $\left(\frac{d_{A}(z_{*})}{D_{V}(z_{BAO})}\right)$. This has been given in \cite{giostri12}. From this one can obtain \cite{giostri12}.

\begin{equation}
\chi^2_{BAO/CMB}={\bf X^tC^{-1}X},
\label{chi2baocmb}
\end{equation}
where
\begin{equation}
{\bf X}=\left(
          \begin{array}{cccc}
          \displaystyle\frac{d_A(z_*)}{D_V(0.106)} -30.95 \\
            \displaystyle\frac{d_A(z_*)}{D_V(0.2)} -17.55 \\
            \displaystyle\frac{d_A(z_*)}{D_V(0.35)} -10.11 \\
             \displaystyle\frac{d_A(z_*)}{D_V(0.44)} -8.44 \\
             \displaystyle\frac{d_A(z_*)}{D_V(0.6)} -6.69 \\
              \displaystyle\frac{d_A(z_*)}{D_V(0.73)} -5.45 \\
          \end{array}
        \right)
\end{equation}
and
\begin{equation}
{\bf C^{-1}}=\left(
          \begin{array}{cccccc}
          0.48435 & -0.101383 &-0.164945 &-0.0305703 &-0.097874 & -0.106738\\
           -0.101383 & 3.2882 & -2.45497 & -0.0787898 & -0.252254 & -0.2751\\
           -0.164945 & -2.45497 & 9.55916 & -0.128187 & -0.410404 & -0.447574\\
           -0.0305703 & -0.0787898 & -0.128187 & 2.78728 & -2.75632 & 1.16437\\
          -0.097874 & -0.252254 & -0.410404 & -2.75632 & 14.9245 & -7.32441 \\
         -0.106738 & -0.2751 & -0.447574 & 1.16437 & -7.32441 & 14.5022  \\
          \end{array}
        \right)
\end{equation}
is the inverse covariance matrix. The correlation coefficients for the  $r_s/D_V$  pair of measurements at $z=(0.2, 0.35)$, $z=(0.44, 0.6)$ and $z=(0.6, 0.73)$, respectively have been given by \cite{Percival,blake11}.

\begin{figure}
\begin{center}
\begin{tabular}{|c|c|}
\hline
 & \\
{\includegraphics[width=2.6in,height=2in,angle=0]{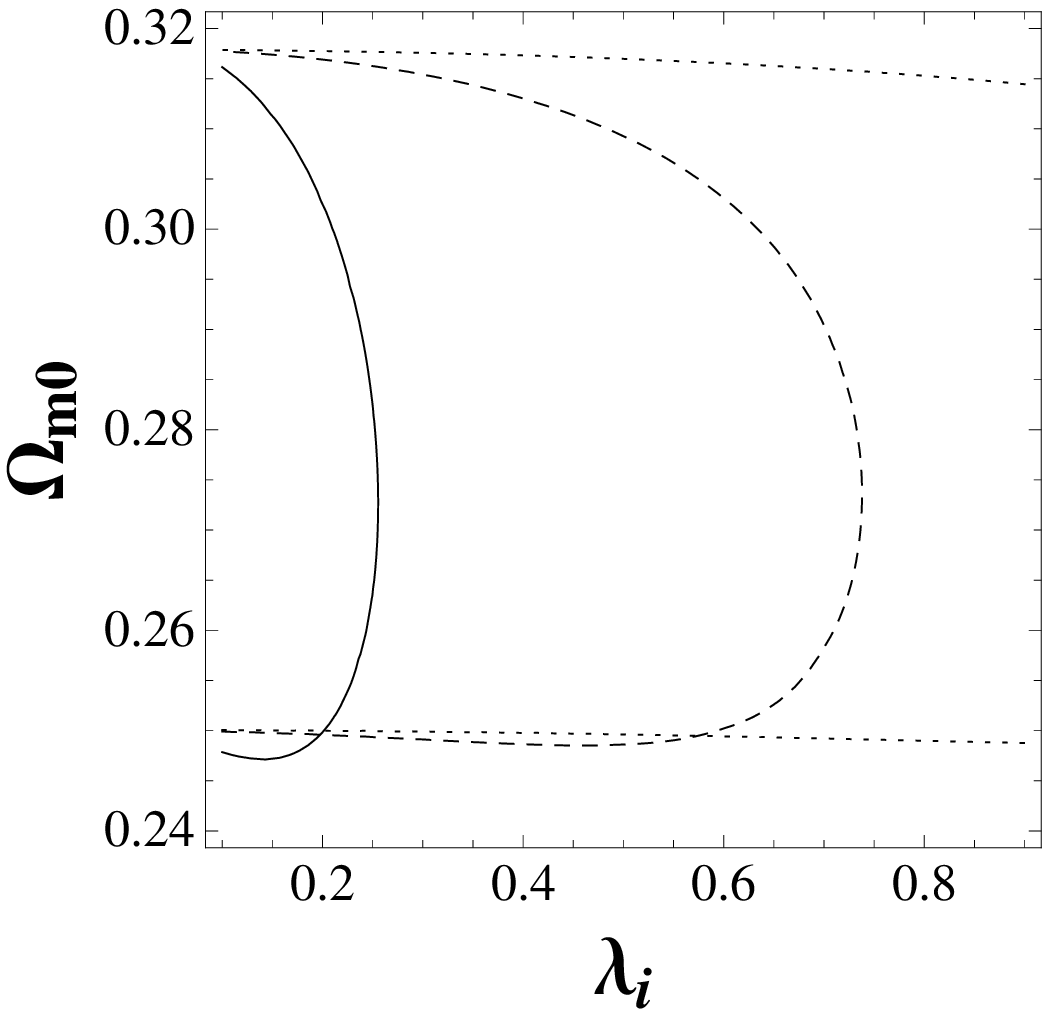}}&
{\includegraphics[width=2.6in,height=2in,angle=0]{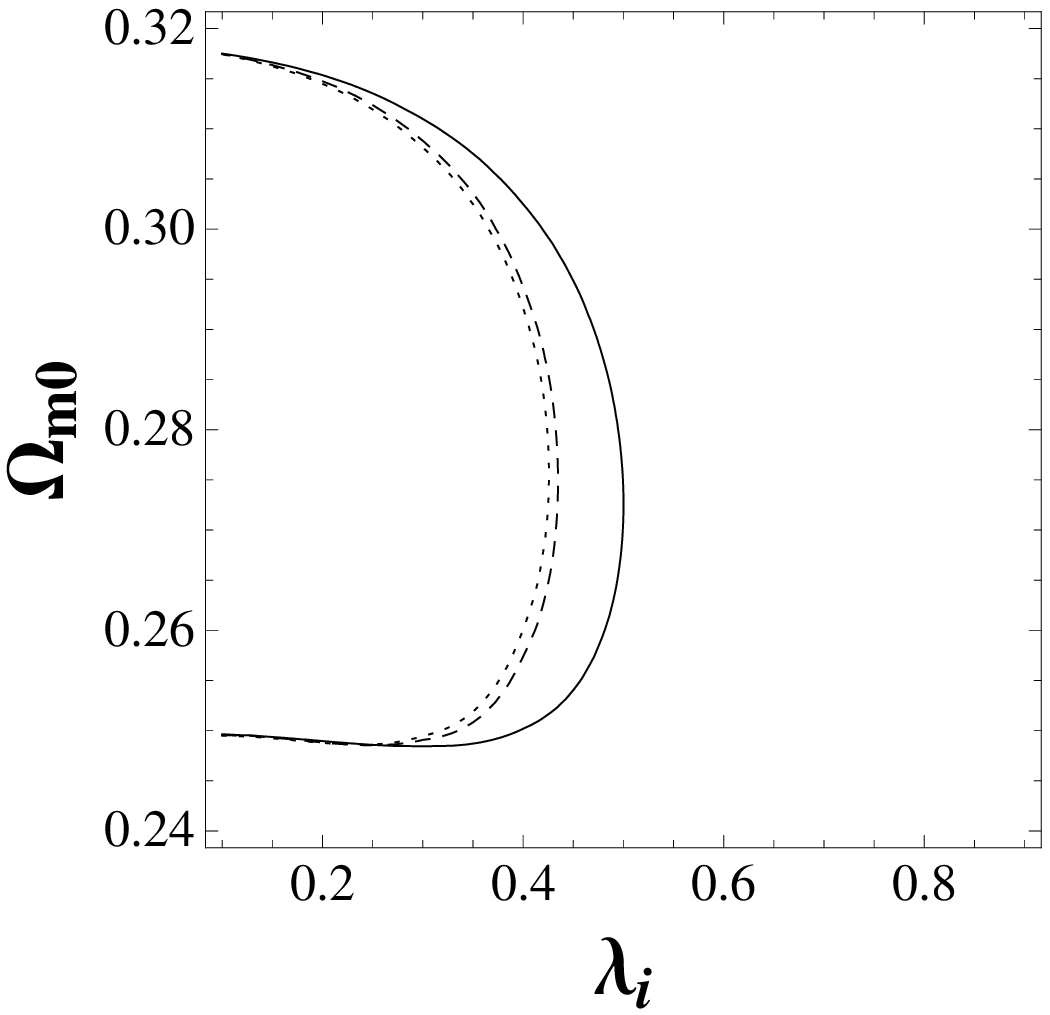}}
\\
\hline
{\includegraphics[width=2.6in,height=2in,angle=0]{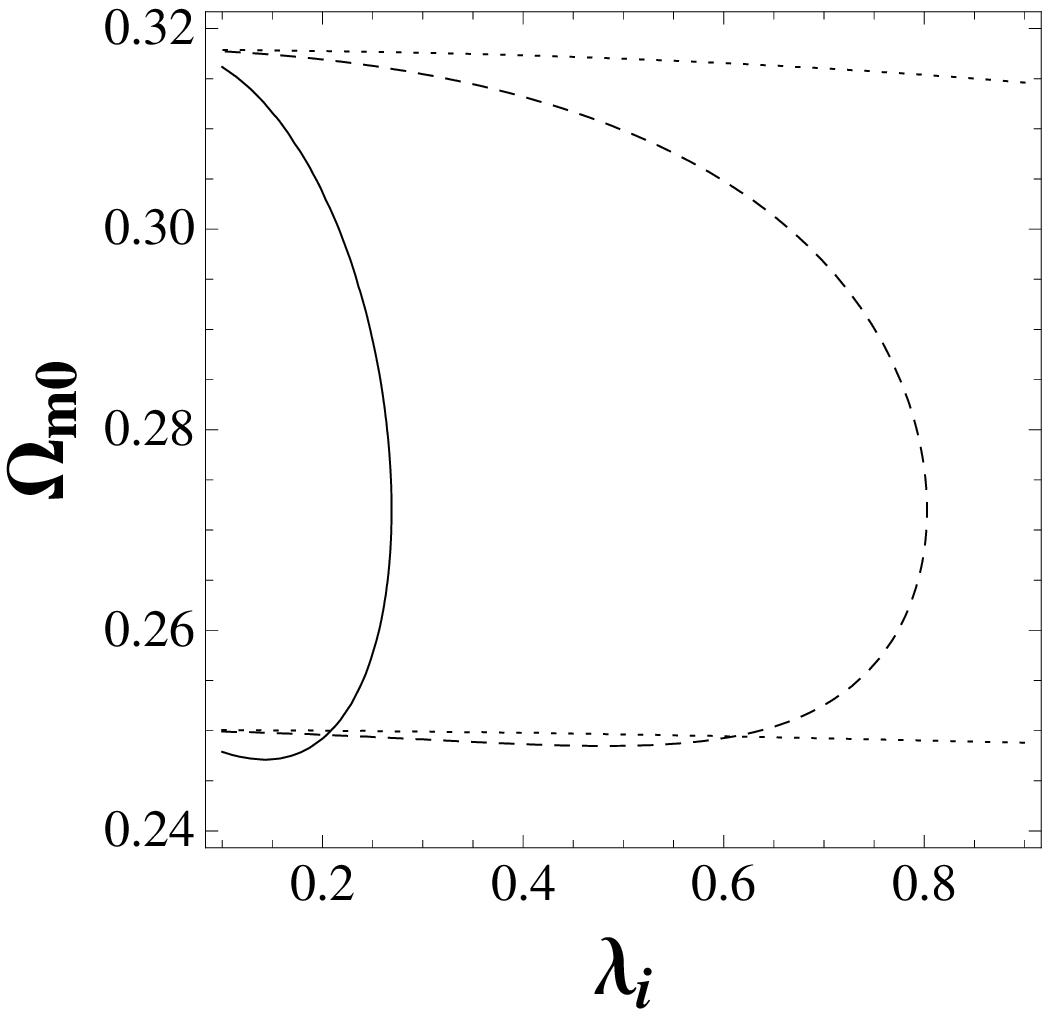}}&
{\includegraphics[width=2.6in,height=2in,angle=0]{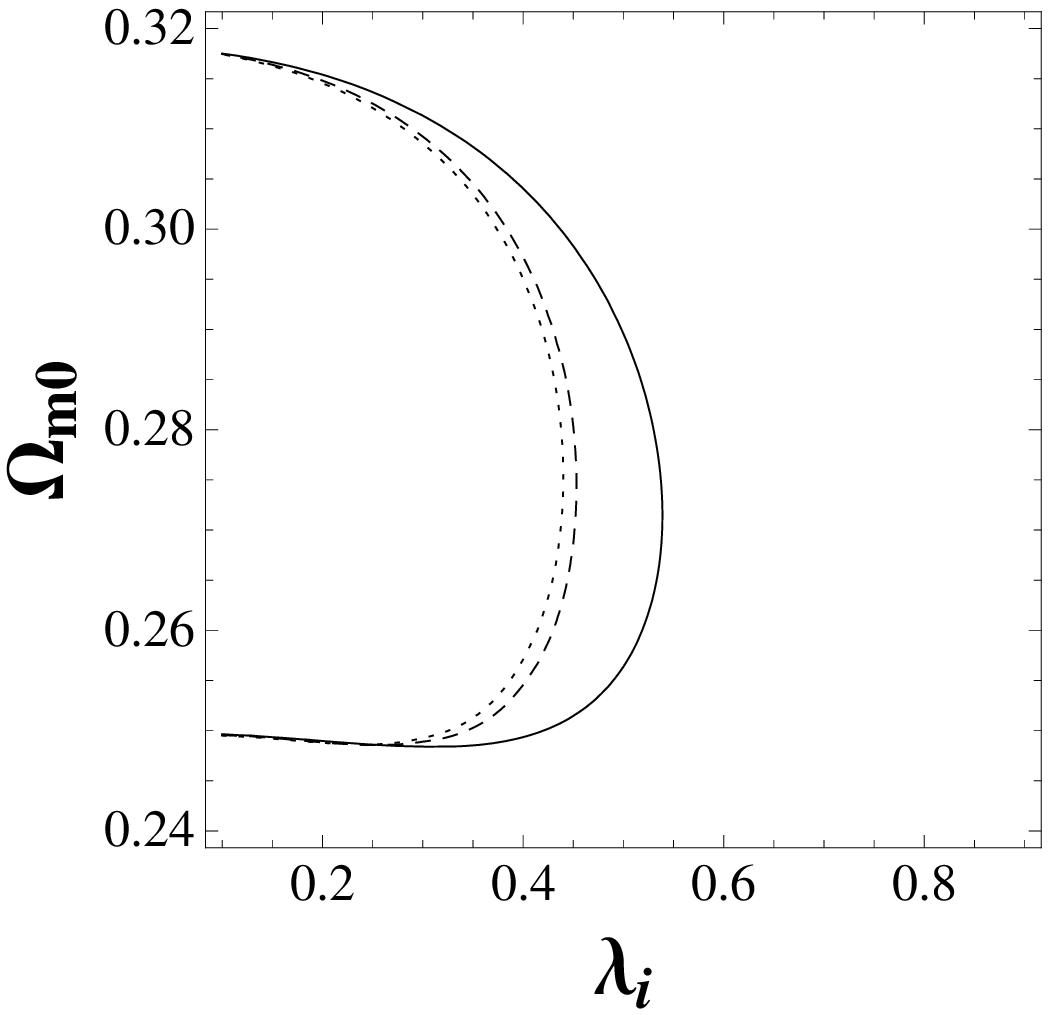}}
\\
\hline
\end{tabular}
\caption{Allowed region in $2\sigma$ cnfidence level in the $\Omega_{m0}-\lambda_{i}$ plane. Top ones are for Linear potential while the bottom ones for quadratic potential. The left ones are for $\alpha=0.1$ (solid), $\alpha=0.5$(dashed) and $\alpha=1$(dotted) keeping $\beta=0.1$. The right ones are for $\beta=0.1$(solid), $\beta=0.5$(dashed) and $\beta=1$(dotted) keeping $\alpha=0.3$. }
\end{center}
\end{figure}

The results are shown in Figure (6) and Figure (7) where we show the $2\sigma$ confidence contours in the $\Omega_{m0}-\lambda_{i}$ plane for different values of $\alpha$ and $\beta$. Dependence on $\alpha$ is such that for smaller values of $\alpha$, there is an upper bound on $\lambda_{i}$ which makes the cosmological evolution very close to the $\Lambda$CDM. But as one increases $\alpha$, this upper bound shifts towards the higher values of $\lambda_{i}$, thereby allowing large deviation from $\Lambda$CDM behaviour. This is true for all the potentials. The allowed deviation from the $\Lambda$CDM behaviour is highest for inverse-squared potential.

On the other hand, if one varies $\beta$ keeping $\alpha$ fixed, there is also an upper bound on $\lambda_{i}$. But as one increases the value of $\beta$, this upper bound shifts towards the smaller value of $\lambda_{i}$ forcing the models to behave very similar to $\Lambda$CDM. This is opposite to the $\alpha$ dependence.  One should note that $\alpha$ and $\beta$ control the contribution from the ${\cal L}_{2}$ and ${\cal L}_{3}$ part of the total action. Hence bigger contribution from ${\cal L}_{2}$ term allows larger deviation from $\Lambda$CDM whereas bigger contribution from ${\cal L}_{3}$ restricts the model to behave more close to $\Lambda$CDM.
\begin{figure}
\begin{center}
\begin{tabular}{|c|c|}
\hline
 & \\
{\includegraphics[width=2.6in,height=2in,angle=0]{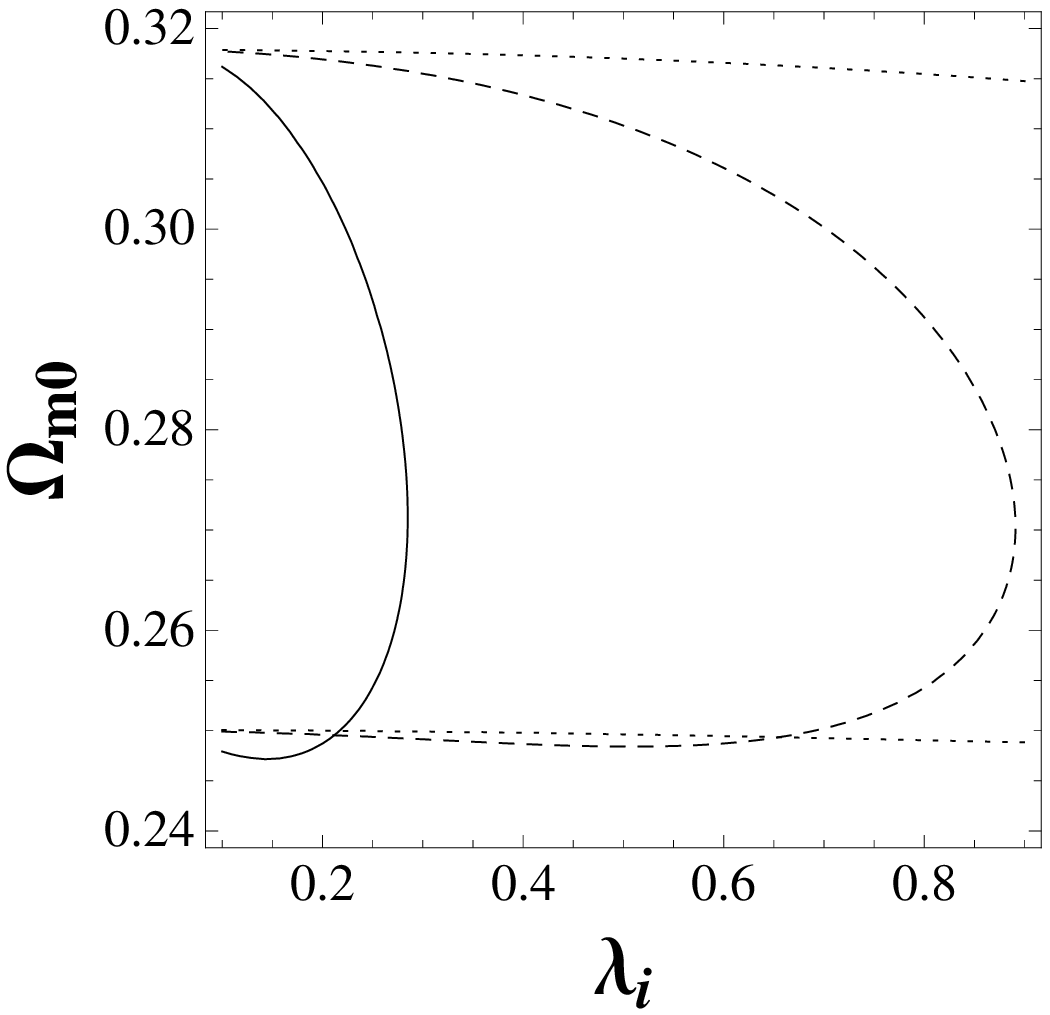}}&
{\includegraphics[width=2.6in,height=2in,angle=0]{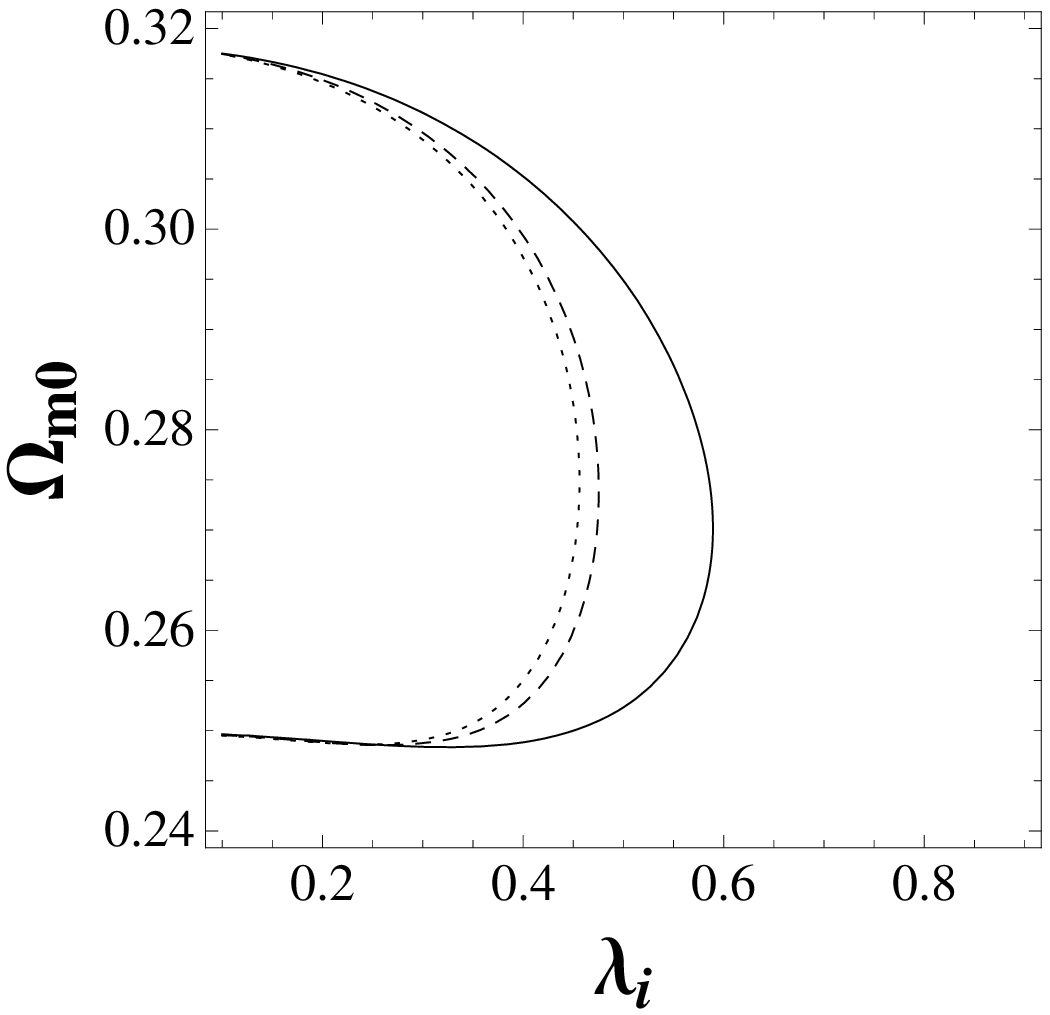}}
\\
\hline
{\includegraphics[width=2.6in,height=2in,angle=0]{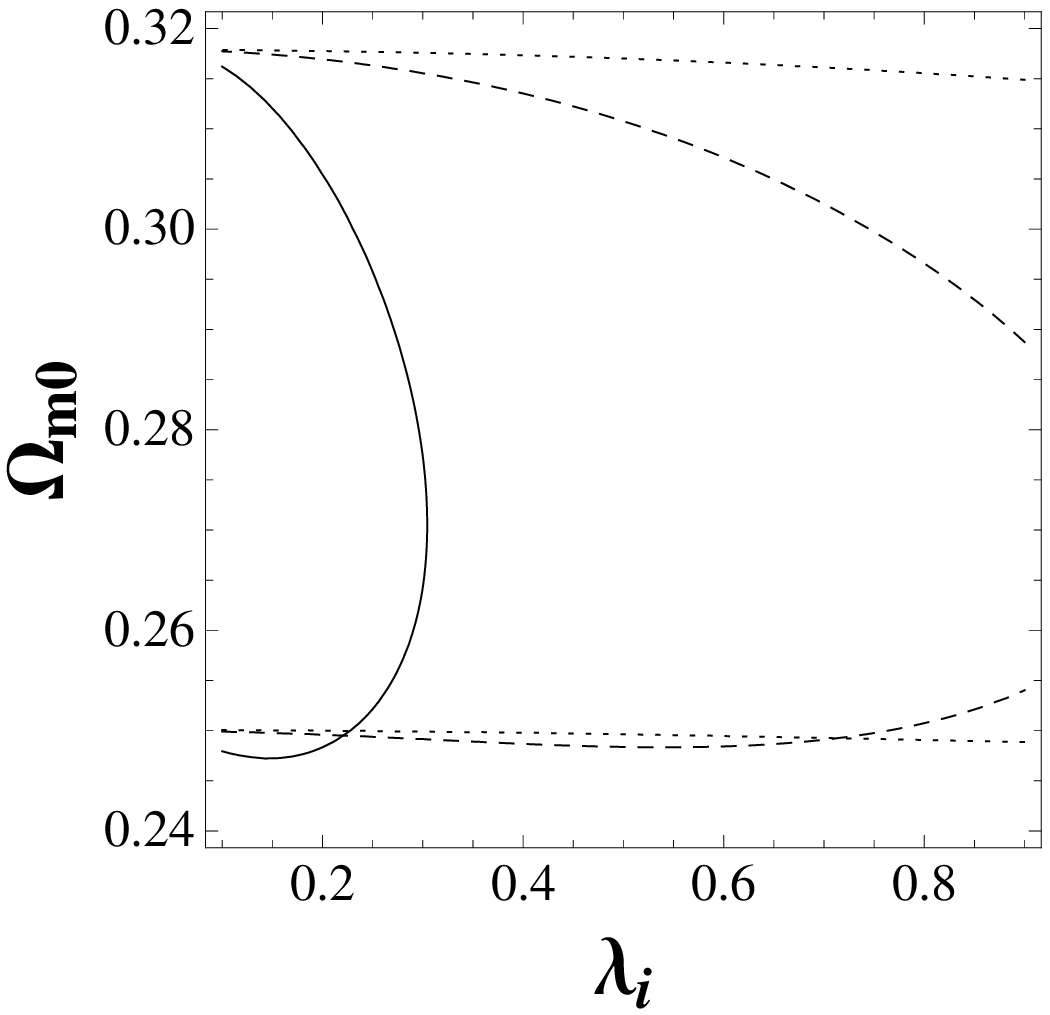}}&
{\includegraphics[width=2.6in,height=2in,angle=0]{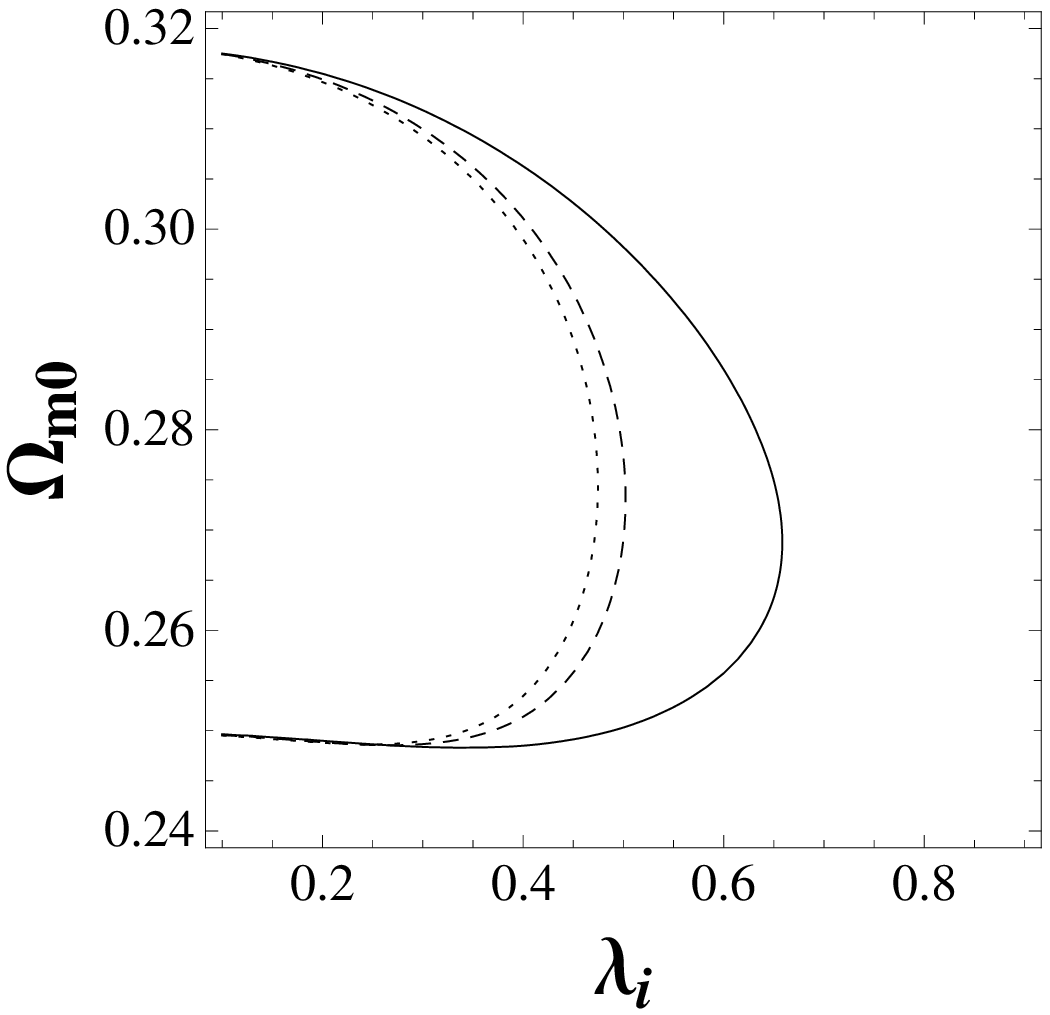}}
\\
\hline
\end{tabular}
\caption{ Same as Figure (6) but the top ones are for exponential potential while the bottom ones are inverse squared potentials.}
\end{center}
\end{figure}

\section{Conclusion}
\vspace{5mm}

In this work, we investigate the late time evolution of the universe in a DBI-Galileon Model where a Minkowski Brane is embedded in a Minkowski bulk. To simplify our analysis, we only keep the first three terms in the total action which under weak-field limit reproduce the standard Galileon terms present in the decoupling limit of DGP model.  We also keep a general potential term $V(\pi)$ in the action instead of standard linear term and study different choices for $V(\pi)$.  We assume the field $\pi$ to be initially frozen due to large Hubble damping and that it behaves like a cosmological constant. However, with time the field slowly thaws out from the frozen state and starts deviating from $w=-1$. The deviation depends on the initial value of te slope of the potential, $\lambda_{i}$. For smaller values of $\lambda_{i}$, the evolution for all the potentials remain close to $\Lambda$CDM throughout. As one increases the value of $\lambda_{i}$, the evolution start deviating from $\Lambda$CDM. We study the degeneracies for the different potentials using the statefinder hierarchies. Finally we constrain our model parameters using the recent observational data. We show that larger contribution from ${\cal L}_{2}$ part results in larger deviation from $\Lambda$CDM behaviour whereas larger contribution contribution from ${\cal L}_{3}$ part restricts the models to behave more close to $\Lambda$CDM. This is true for all the potentials.

Although we have not studied the complete action for the DBI-Galileon model, still this study gives some interesting observational consequences for first three terms of the full action. It will be worthwhile to study the observational consequences with full DBI-Galileon action and this will be our future aim.

\section{Acknowledgment}
The author SPA is fully supported by the SERC, Dept of Science and Technology, Govt of India through the grant SR/S2/HEP-43/2009. The author AAS is partially supported by the same grant.

\end{document}